\documentclass[journal]{aa}
\usepackage{graphics}

\newcommand{\gr}{$\gamma$-ray \,}
\newcommand{\grs}{$\gamma$-rays \,}
\begin{document}

\thesaurus{02.18.5; 09.03.2; 10.07.1; 13.07.3}

\title{Broad-band diffuse gamma ray emission of the Galactic Disk}
\author{
F.A. Aharonian\inst{1},
A.M. Atoyan \inst{2},
 }
 \institute{Max Planck Institut f\"ur Kernphysik,
Postfach 103980, D-69029 Heidelberg, Germany \and
 Yerevan Physics Institute, Alikhanian Br. 2, 375036 Yerevan,
 Armenia}

 \offprints{}
 \date{Received      ; accepted  }
 \authorrunning{F.A. Aharonian and A.M. Atoyan}
 \titlerunning {Diffuse galactic gamma-ray emission}
 \maketitle

\begin{abstract}
The  contributions of different  radiation  mechanisms 
to  the diffuse  \gr emission of the galactic disk are studied  
in a broad energy region  from $10^4$ to  $10^{14}$  eV. 
Our analysis shows that at energies between 1 and 100 MeV 
the  radiation is dominated  by  the bremsstrahlung of 
relatively  low energy, typically  less than  1 GeV  electrons, but 
with a non-negligible contribution  from the inverse Compton (IC)  
scattering of higher energy  electrons.  Also, a significant  
fraction  of the radiation observed at  energies  around 1 MeV  could be 
contributed  by mildly relativistic   positrons annihilating ``in flight''  
with the ambient thermal  electrons.  At energies from  100 MeV to 100 GeV 
the $\gamma$-ray flux is dominated by interactions of cosmic ray protons 
and nuclei with the  ambient gas through production and subsequent 
decay of secondary $\pi^0$-mesons.  The interpretation of  the GeV 
$\gamma$-ray emission  of  the inner  Galaxy as a truly diffuse radiation  
requires a substantially  harder spectrum of relativistic  protons and nuclei 
in the interstellar medium  
compared  with the local cosmic ray spectrum measured directly in the solar 
neighborhood.  In the very high energy  
domain, $E_\gamma \geq  100 \, \rm GeV$, the contribution of the IC component 
of radiation may become comparable with, or even could exceed the fluxes of 
$\pi^0$-decay component  if  the energy spectrum of electrons injected 
into the interstellar medium extends well beyond  1 TeV.
Another signature of  multi-TeV electrons is the synchrotron radiation  
which could account for a  significant  fraction of  the diffuse hard 
X-ray flux of the galactic ridge.
The future detailed studies of  spatial and  spectral characteristics of  
\gr emission of the galactic disk, especially  at very high energies around
 100 GeV by GLAST, and hopefully also at TeV energies by planned 
ground-based instruments
should  provide  important insight into  the understanding of the 
sites and mechanisms of acceleration of galactic cosmic rays and the  
character of their propagation  in the interstellar magnetic fields. 
\end{abstract}

\section{Introduction}

Diffuse $\gamma$-ray emission of the galactic disk carries unique 
information about the fluxes  and the spatial  distribution
of galactic cosmic rays 
(CRs). Therefore  it is believed that the solution 
of  the long-standing problem of 
the origin of galactic CRs essentially  depends on the  success  
of observational gamma-ray astronomy.
Indeed, the separation of the radiation components associated
with the electronic and nucleonic components in a broad energy region 
from 1 MeV to 100 TeV would allow  determination of the fluxes and energy 
spectra of CRs in different parts of the galactic disk, and thus 
would provide an important  insight to
the  character of propagation of 
CRs  in  the interstellar medium  (ISM). A proper understanding of the 
latter is a necessary  condition for accurate estimates of the luminosity
of the Galaxy in  CRs.   Our present knowledge about the propagation 
of CRs  is based on  conclusions derived from the interpretation of the mass 
composition and the content of the secondary anti-particles 
(positrons, antiprotons) of the {\it locally}  observed CRs.  
Although rather effective  (see e.g. Swordy 1993, 
Strong et al. 1998), this method requires 
a number of  {\it model-dependent} assumptions. 
Moreover, it is not yet obvious  that the locally observed CRs 
could be taken as  undisputed representatives of the whole galactic 
population of relativistic particles. For example, 
Erlykin et al. (1998) recently argued that the fluxes of CRs 
could be  dominated by a single or few
local sources/accelerators. This statement is certainly true at 
least for the observed $\geq 1 \, \rm TeV$ electrons which suffer 
severe synchrotron and IC energy losses,
and thus could reach us, for any reasonable diffusion coefficient, 
only from the sites no farther than a few hundred  
parsecs (Nishimura et al. 1980, Aharonian et al. 1995).  

Therefore the diffuse galactic \grs  seem to be the  best carriers 
of information about the production sites and propagation of 
accelerated charged particles in the galactic disk  
(see e.g. Ramana Murthy \& Wolfendale 1993). It should be noted that
the diffuse non-thermal synchrotron radiation of the 
ISM at radio  and possibly also at X-ray wavelengths  provide an additional
and  complementary information, but it concerns only 
the {\it electronic} component of  CRs in two extreme  energy bands 
below  $1 \, \rm GeV$ and above  $100 \, \rm TeV$, respectively.

The extraction  of the  truly   diffuse \gr emission, 
i.e. the radiation  produced by CR electrons,  protons and 
nuclei interacting with the ambient interstellar gas and  photon 
fields, is not an easy task  
because of a non-negligible contamination due to    
weak  but  numerous  
unresolved discrete sources. Before the launch of the 
Compton Gamma Ray Observatory (GRO)  the information about the 
diffuse galactic \gr background 
was essentially  limited to the energy region between 100 MeV and few
GeV
obtained by the SAS-2 (Fichtel et al. 1975) and COS B 
(Mayer-Hasselwander et al. 1982)  $\gamma$-ray missions. The 
observations and theoretical models based on 
the results of these satellites  were comprehensively 
reviewed by Bloemen (1989).
In brief, these data  have revealed
a good correlation between the
high energy \gr fluxes and the column density of the interstellar
hydrogen which was a  demonstration  of 
the existence of a truly diffuse galactic gamma radiation.

The observations   of the diffuse \gr background conducted in 90's 
by the  OSSE, COMPTEL, and EGRET  detectors aboard Compton GRO 
resulted in  good quality data over five decades in energy of 
$\gamma$-rays (see Hunter et al. 1997a, and references therein). 
These results initiated 
extensive theoretical studies of different  \gr production  mechanisms 
in the ISM  (e.g. Bertsch et al. 1993, Giller et al. 1995, 
Fathoohi  et al. 1995, Gralewicz et al. 1997, 
Mori 1997, Porter \& Protheroe 1997,  Moskalenko \& Strong 1998,
Strong et al. 1998,  Pohl \& Esposito 1998).

In this paper we report  the results of our study 
of different \gr production processes  in the ISM.   
We discuss the fluxes of diffuse non-thermal galactic radiation 
of both nucleonic and electronic origin in a very broad energy 
region from  hard X-rays  to ultra-high energy $\gamma$-rays.
Although at the first glance the problem seems to be very   
complicated and confused because of several competing  
production mechanisms,  the existing data of diffuse galactic 
$\gamma$-radiation  do allow rather definite conclusions concerning the 
relative contributions  of different  production processes 
in each specific energy band of $\gamma$-rays. A separate interest represents 
the  diffuse $\gamma$-radiation in the very  high energy 
(VHE)  domain.   In this paper we limit our  study  by 
the diffuse radiation of the {\it inner part} 
of Galaxy  at $315^\circ \leq l \leq  45^{\circ}$. The inner Galaxy  
is not only the 
experimentally best studied region in diffuse $\gamma$-rays, but 
also, it  presents a prime interest because 
of an enhanced, as currently believed, spatial concentration
of CR sources in the central part  of the Galaxy.
We emphasize the importance of the 
multiwavelength approach in solution of the problem, 
and predict  a range of  \gr fluxes  which could be 
examined  by forthcoming  satellite-borne and ground based \gr detectors.

\section{The spectra of cosmic rays in the inner Galaxy} 

We consider a model that assumes diffusive 
propagation of relativistic protons/nuclei
and electrons in the  Galaxy. We will compare our model calculations 
with observations of the diffuse $\gamma$-radiation of the 
galactic disc  at  $|b| \leq 5^\circ$.  Thus, although
the halo of galactic CRs may extend up to heights of a few kpc (e.g. see
Berezinsky et al. 1990, Bloemen et al. 1993), we will need to know 
the {\it mean} spectrum of CRs only in a region close to the galactic       
plane. We approximate this region as a disk 
with a half-thickness $h\simeq 1\,\rm kpc$, and a
surface 
$S_{\rm tot}\approx 2 S_{\rm G}$, where $S_{\rm G}= \pi R_{\rm G}^2$, 
and $R_{\rm G}\sim 15 \,\rm kpc$ is the mean radius of the Galaxy.  

An important parameter
for calculations of the diffuse $\gamma$-radiation is the mean
line-of-sight depth $l_{\rm d}$ in the direction of the inner Galaxy, which 
basically represents the central part of the galactic disk, with
some radius $R_{\rm d}\leq R_{\odot}$, where 
$R_{\odot}\simeq 8.5 \,\rm kpc$ is the distance of the Sun from the
center of the Galaxy. Because the densities of
the gas and photon fields  in the inner Galaxy are
generally estimated to be significantly higher than at galactocentric 
distances $R\geq R_{\odot}$, the inner galactic disk should
be responsible for
most of the diffuse flux detected at low galactic latitudes,
neglecting even the effect of a possible gradient of CR density towards
the center of the Galaxy. Thus, the mean value of $l_{\rm d}$ can be 
reasonably estimated as $\sim 15 \,\rm kpc$. 

The diffusion equation for the energy distribution  $f \equiv 
f({\bf r},E,t)$ of relativistic particles can be written in a general
form as (Ginzburg \& Syrovatskii 1964):
\begin{eqnarray}
\frac{\partial f}{\partial t} & = & {\rm div}_{\bf r} 
(D\,{\rm grad_{\bf r}} f) 
- {\rm div}_{\bf r}({\bf u} f) + \nonumber \\
& &
 \frac{\partial}{\partial E}(P f) + A[f] \; ,
\end{eqnarray}
where  ${\bf u} \equiv {\bf u}({\bf r})$ is the fluid velocity of the 
gas containing relativistic particles, and  
$P\equiv P({\bf r}, E)= -{\rm d} E/{\rm d} t$
describes their total energy losses  
where we include also the adiabatic energy loss 
term $P_{\rm adb} = {\rm div \bf u} E/3$ (e.g. Owens \& Jokipii 1977, 
Lerche \& Schlickeiser 1982).
$D\equiv D({\bf r}, E)$  is the spatial  diffusion coefficient, and  
$A[f]$ is a functional standing for various acceleration terms of
relativistic particles (i.e. the sources of CRs).

The  integration of Eq.(1) over the volume $V=2h S_{\rm G}$  results in
a convenient equation for the total energy distribution function of 
particles $N(E,t)={\int} f \,{\rm d}^3 r\,$ in the Galactic disk 
at the heights $|z| \leq h$. The term $\partial f / \partial t$  leads
to $\partial N / \partial t$. 
The volume integral of the two first terms in the right hand side of Eq.(1)  
results in:
\begin{eqnarray}
\int_{ V} [ {\rm div}_{\bf r} (D \,
{\rm grad_{\bf r}} f)  -  {\rm div}_{\bf r}({\bf u} f) ]\,{\rm d}^3 r 
& = &  \\ 
  \oint_{S_{\rm tot}} 
D\,({\bf e}\, {\rm grad}_{\bf r} f)\, {\rm d} s 
& - & \oint_{S_{\rm tot}} ({\bf e \, u}) f  \, {\rm d} s  \; . \nonumber
\end{eqnarray}
Here {\bf e} is a unit vector perpendicular to the 
surface element ${\rm d}s$  directed outward from the disk. 
These terms describe the diffusive and convective escape of 
particles from the disk through its 
surface $S_{\rm tot}\approx 2S_{\rm G}$. 

The first surface integral in Eq.(2) can be simplified if we  
take into account that the CR density at heights
$z\leq h \simeq 1\,\rm kpc$,  where the diffusion dominates (see below),
may be significantly higher than at $z > h$, 
and approximate $({\bf e} \,{\rm grad} f) \simeq - n(E,t)/\Delta x$,
where  $n(E,t)\equiv \bar{f}(E,t)$ is the volume averaged energy 
distribution function of relativistic particles, 
and $\Delta x \equiv \Delta x (E)\,$ is the 
characteristic  thickness of the transition layer  
describing the decline (i.e. the gradient) of the density of particles
at energy $E$. Taking into account that the 
total energy distribution of particles $N(E,t)=2 h S_{\rm G}\,n(E,t) $,  
one finds 
\begin{equation}
\oint_{S_{\rm tot}} D\,({\bf e} \, {\rm grad} f ) = 
- \frac{n \, \overline{D} \, 2 S_{\rm G} }{\Delta x }
  = - \frac{N}{\tau_{\rm dif}}  ,
\end{equation}
where $\tau_{\rm dif} $ has  a meaning of a  characteristic diffusive escape 
time of relativistic electrons  from the Galactic disk:  
\begin{equation}
\frac{1}{\tau_{\rm dif}(E)} =
 \frac{\overline{D}(E)}{h \Delta x(E)}\; ,
\end{equation}
Here $\overline{D}(E)$ corresponds to the
mean diffusion coefficient ${D}_{\rm r}({\bf r},E)$ 
on the surface $S_{\rm G}$ of the disk.

The second surface integral in Eq.(2) can be reduced to the 
form $- N(E,t)/\tau_{\rm conv}$, which describes 
a convective escape of electrons from the disk through its surface  
due to the galactic wind driven by the
pressure of CRs and of the thermal gas 
(e.g. Bloemen et al. 1993, Breitschwerdt et al. 1993, Zirakashvili et al 1996):  
\begin{equation}
\tau_{\rm conv} \simeq  a h/{ u}\; .
\end{equation} 
Here $u$ is the wind velocity on the surface of the Galactic disk
which could reach $\sim 50 \,\rm km/s$ (Zirakashvili et al. 1996) at 
the height   
$h=1\,\rm kpc$. Thus, 
the {\it mean} convective escape time of CRs from the Galactic disk 
can be estimated as $\tau_{\rm conv} \simeq 2 \times 10^7 a \, \rm yr$

The parameter $a $ in Eq.(5) is the ratio of the mean density 
$n(E)$ of particles in the disk  ( i.e. an average over $-h
\leq z \leq h$ )   to their density
$f(h,E)$ at the disk surface, $z=\pm h$.  
Therefore one could expect that  $a \geq 1 $ 
and generally it may be energy-dependent as well, $a= a(E)$.
Calculations of the spatial and energy distribution $f(z,E)$ of the galactic 
CRs in the framework of the diffusion-convection model show 
(Lerche \& Schlickeiser 1980; see also Lerche \& Schlickeiser 1982, 
Bloemen et al. 1993) that at elevations $z \ll z_{\rm c}(E)$,
where $z_{\rm c}(E)$ is a characteristic height of the diffusion dominated
region for particles with energy $E$, the spatial density $f(z, E)$ is 
almost independent of $z$. Thus, at energies $E \geq E_{\ast}$, where 
$E_{\ast}$ is found from the equation $z_{\rm c}(E) = h$, the parameter 
$a\sim 1$. In the approximation $u(z) = v_0 z$ for the galactic wind speed
and $D(E) \propto E^{\delta_1}$ with $\delta_1 \leq 1$ for the diffusion 
coefficient, the height $z_{\rm c} = \sqrt{2 D(E) / v_0 (1+\delta_1/6)}$
(see e.g. Bloemen et al. 1993).  This is basically the height at which 
the characteristic time scale of the diffusive propagation $\sim z^2/D$ 
equals the convection time scale $v_{0}^{-1}$.  The energy $E_{\ast}$ can
be estimated of order of a few GeV, taking into account that at these 
energies $D\sim 10^{28}\,\rm cm^2/s$ (e.g. Berezinsky et al. 1990)
and that $v_0 \sim 50 \,\rm km/s\,kpc$ (e.g. Zirakashvili et al. 
1996), which result in $z_{\rm c} \simeq 1\,\rm kpc$. 

In the convection dominated region, $z \gg z_{\rm c}(E)$, the spatial 
density of particles starts to decline as $f(z, E) \propto 
(z/z_{\rm c})^{- \kappa}$ with $\kappa \simeq 1.3-1.4$ (see Lerche \&
Schlickeiser 1992, Bloemen et al. 1993), thus  $f(h, E)/f(0, E) 
\propto E^{-\kappa \delta_{1}/2}$. Therefore 
at energies $E\ll E_{\ast}$ the parameter $a$ should gradually increase,
and in principle could be approximated as 
$a(E) \propto (E/E_{\ast})^{- \lambda}$, with an exponent $\lambda$
much smaller than $\kappa \delta_1/2$ because at these energies
the mean particle density $n(E)$, as compared with $f(0,E)$, should 
also decline. However, taking into account that 
$\lambda$ is significantly less than 1, and in order to avoid an 
introduction of an additional model 
parameter, below we approximate the convective 
escape time as energy-independent, i.e. 
with $a(E)  \sim 1$ at all energies. For the CR proton component this 
approximation is well justified, because for $E_\ast$ of order of a few
GeV the energy region  $E\ll E_{\ast}$ effectively corresponds to  
sub-relativistic protons which do not represent an interest for this study.
Such a simplification appears   
reasonable also for CR electrons, 
because at energies $E\ll 1\,\rm GeV$ the spectral 
modifications of the electrons are defined mainly by their Coulomb energy 
losses which take place  on time scales significantly shorter than the escape 
losses.

The volume integral of the third term on the right side of Eq.(1)
describes the volume-averaged energy losses of the electrons
with the rate $\overline{P}(E)$.   
At last, the volume integral
of the 4th term in Eq.(1) describes the sources of accelerated
particles
$Q(E,t)$ inside the 
volume $V$. The final equation for the overall distribution
of electrons then reads:
\begin{equation}
\frac{\partial N}{\partial t} = \frac{\partial (\,\overline{P}\,N\,)}
{\partial E} \, - \, \frac{N}{\tau_{\rm esc}}\,  
 +\, Q \; .
\end{equation}
Here $\tau_{\rm esc}$ is the   
"diffusive + convective" escape time of particles: 
\begin{equation}
 \tau_{\rm esc}(E) = \left[ \frac{1}{\tau_{\rm dif}(E)}
+\frac{1}{\tau_{\rm conv}}\right]^{-1}\; .
\end{equation}

The range of actual energy dependence of 
$\tau_{\rm esc}$ is limited at high energies because    
the diffusive escape time 
cannot be less than the light travel time $ h / c$. 
This obvious requirement formally 
follows from the condition that the characteristic 
length-scale $\Delta x(E)$ for spatial gradients in the 
distribution function $f(z, E)$ 
cannot be less than the mean electron scattering path 
$\lambda_{\rm sc}(E)$, otherwise the diffusion approximation 
implied in Eq.(1) fails. 
Taking into account that for relativistic particles $D \sim \lambda_{\rm
sc} c /3$, from Eq.(4) we find that indeed 
$\tau_{\rm dif}(E) \geq \tau_{\rm min} \simeq 3 h/c$.
The diffusive escape time can be  then presented in the form  
\begin{equation}
\tau_{\rm dif}(E) = \tau_{10} \,(E/E_{10})^{- \delta}
\, +\, \tau_{\rm min}\; ,
\end{equation}
where  $E_{10}\equiv 10\,\rm GeV$. 
Since $\tau_{\rm dif}$  increases  
for decreasing $E$,  $\tau_{\rm esc}$  given by Eq.(7) becomes
energy {\it independent} below some
$E_{\ast}$ when $\tau_{\rm dif}(E_{\ast}) = \tau_{\rm conv}$.
 Neglecting at these energies $\tau_{\rm min}$ in
Eq.(8), from Eq.(7) then follows that in the case of a pure power-law approximation
for $\tau_{\rm dif}(E)$ as in Eq.(8)  
the overall escape time can be presented
in the form $\tau_{\rm esc}\approx \tau_{\rm conv}/[1+(E/E_{\ast})^\delta]$.

Note that generally the power-law index 
$\delta$ in Eq.(8) for the diffusive escape time in the Leaky-box type 
Eq.~(6)  should be smaller than the  index $\delta_1$ of 
the diffusion coefficient $D(E) \propto E^{\delta_1}$.
Since a faster  diffusion of more energetic particles tends to 
smooth out the gradients of the distribution function 
$f(z, E)$ more effectively, the characteristic 
length-scale $\Delta x(E)$ would increase with energy. 
For a power-law approximation $\Delta x(E)
\propto E^{\,\delta_2}$ the index $\delta_2 >0$, therefore 
from Eq.(4) follows that $\delta = \delta_1 - \delta_2 < \delta_1 $.   
This consideration may help to qualitatively understand a formal
discrepancy between CR spectral modifications due to particle
propagation effects predicted in the framework 
of simplified Leaky-box models and more accurate diffusion-convection
propagation models. 
Calculations for the latter models show (e.g. Lerche \& Schlickeiser
1980, Bloemen et al. 1993) that at elevations $z\ll z_{\rm c}(E)$,
where the diffusive propagation dominates over convection,
the initial power-law
spectrum of injected particles, $Q(E)\propto E^{-\Gamma_0}$,
 steepens by a factor $\propto E^{-\delta_1}$, whereas
at $z \gg z_{\rm c}$ the increase of the effectiveness of the
convective propagation results (in the case of negligible
energy losses) in a steepening to only  a half of the power-law 
exponent of the diffusion
coefficient, $f(z,E)\propto E^{-\Gamma_0 -\delta_1 /2}$. Meanwhile, 
the (diffusive) escape
losses in the Leaky-box models result in a single power-law spectrum 
$n(E) \propto E^{-\Gamma_0 -\delta}$. Such a  
difference between the predictions of the CR spectra 
in the framework of diffusion-convection models and  
Leaky-box type models 
could be qualitatively explained, if we take 
into account that $n(E)$ represents the mean particle spectrum
integrated over $z\leq h$, which is therefore contributed  
(in fractions varying with $E$) by both
`diffusion' and `convection' dominated regions. Thus,
the power-law index $\delta$ for the escape time should be effectively in 
the region $\delta_1/2 < \delta < \delta_1$.

Assuming a time-dependent injection function $Q(E,t)$, 
Eq.(6) can be used for determination of 
the overall energy distribution of relativistic particles in a general
case of a non-stationary source.
If the energy losses are independent of time,  the 
solution to this equation, in terms of spatial density functions 
$n=N/V$ and $q=Q/V$, reads:
 \begin{eqnarray}
n(E,t) & = & \frac{1}{P(E)} \int_{0}^{t}
P(\zeta_t) q(\zeta_t,t_1) \times \nonumber \\
& & \exp \left( -\int_{t_1}^{t}\frac{{\rm d} x}
{\tau_{\rm esc}(\zeta_x)}\right)
{\rm d} t_1\, .
\end{eqnarray}
Here 
the  variable $\zeta_t$ corresponds to the energy of a  particle at 
an instant $t_1\leq t$ which has the energy $E$ at the 
instant $t$, and is determined from the equation
\begin{equation}
t-t_1 =\int_{E_{\rm e}}^{\zeta_t}{\frac{{\rm d}E_1}{P(E_1)}} \, .
\end{equation}
For a quasi-stationary injection of electrons into the ISM 
on time-scales exceeding the
 escape time $\tau_{\rm esc}(E)$
 the energy distribution of particles becomes time-independent.

For calculations of the energy distribution of CR protons in the Galaxy 
we take into account the  energy losses connected with their 
inelastic interactions with the ISM gas,   
and the adiabatic losses  of particles in a gradually accelerating
wind. In a simple approximation $u(z)\propto z$, used e.g. by Lerche \&
Schlickeiser (1982) and  Bloemen et al. (1993),
the mean adiabatic energy loss term  $\overline{P}_{\rm adb} (E)$
is found after a simple integration 
$$ 2 S_{\rm G} \int_{0}^{h}\frac{1}{3} \frac{\partial u}{\partial z}
f(z,E) \,{\rm d} z  \simeq N(E) E \frac{u(h)}{3h}$$
in the volume of the disk with $|z|\leq h$. This expression corresponds
to $\overline{P}_{\rm adb}=E/\tau_{\rm adb}$ with $\tau_{\rm adb}=
3h/u(h)\simeq 6\times 10^7 \,\rm yr$ for $u(h=1\,\rm kpc)\simeq 50\,\rm
km/s$.

For the energy losses of particles due to their interactions with the
gas,  which contribute to the overall energy loss term in
Eq.(6), 
we should use the volume-averaged gas density 
$\overline{n}_{\rm H}=\int_{0}^{h} n_{\rm
H}(z) {\rm d} z /h$, where $n_{\rm H} = n_{\rm HI } + 2\times  n_{\rm
H2}$ is the gas density in terms of `H-atoms'. In order to
estimate $\overline{n}_{\rm H}$,  
the  H{\sc I} density distribution by Dickey \& Lockman
(1990) can be taken: a sum of two Gaussians with central densities
$\simeq 0.4$ and $\simeq 0.11 \,\rm cm^{-3}$ and FWHMs of 210 and 
530\,pc,  respectively,  and an exponential 
with the central density $0.064 \,\rm
cm^{-3}$ and a scale height $\simeq 400\,\rm pc$. The molecular gas
layer can be approximated by a further Gaussian with the mid-plane
density $0.3 \,\rm H_2 /cm^3$ and dispersion 
70 pc (Bloemen 1987).  Such a gas density profile 
results in the mean  hydrogen density
$\overline{n}_{\rm H}\simeq 0.15 \,\rm cm^{-3}$ in the region $z\leq
1\,\rm kpc$. 

An essential process which defines the spectra of
both CR protons and electrons in the Galaxy is their energy
dependent escape. 
Beside this, for calculations of the energy distribution 
of the electron component of CRs we
take into account the ionization (Coulomb)
losses which dominate the  overall energy losses of CR electrons in the
ISM at energies 
below a few 100 MeV, the adiabatic losses, the radiative (synchrotron and
inverse Compton) losses, and the bremsstrahlung losses.
Note that the bremsstrahlung loss term has 
practically the same energy dependence as the
adiabatic loss term, $\overline{P}_{\rm brem} \simeq E \tau_{\rm
brem}$ (neglecting a weak logarithmic increase with energy, 
e.g. see Ginzburg 1979), but 
for $\overline{n}_{\rm H}\simeq 0.15\,\rm cm^{-3}$ the cooling time
$\tau_{\rm brem}\sim 3\times 10^{7}/\overline{n}_{\rm H} \,\rm yr$ 
exceeds the adiabatic cooling time by a factor $\sim 3$. Therefore 
for the formation of the energy spectra of CR electrons (but {\it
not} for the radiation flux!) the bremsstrahlung losses are by a factor 
of 3 less effective than the adiabatic losses. 
At energies above several GeV the total energy losses of the electrons 
are dominated by the radiative 
energy losses due to synchrotron emission in the ISM
magnetic field of order of several $\mu \rm G$, and due to the 
IC scattering of the electrons on different diffuse target photon
fields.

Besides these processes, which should be taken into
account for calculations of energy  distribution of CR electrons
in the disk, in this paper we discuss also a possible contribution to the 
fluxes of diffuse galactic radiation caused by  annihilation of
relativistic positrons in `flight' with the ambient thermal electrons. 
In principle, it is possible to include
also this process into calculations of $n(E)$, 
considering the energy distributions for $e^+$ and $e^-$ separately, 
and introducing in the equation for $n_{+}(E)$ an additional term which 
describes the disappearance  of the positrons
due to annihilation on a timescale $\tau_{\rm ann}(E)$. However, this term
does  not have a  significant impact on the formation of the 
spectrum $n_{+}(E)$, because at any energy  the annihilation time 
$\tau_{\rm ann}(E)=(\sigma_{\rm ann}(E) v \overline{n}_{\rm H})^{-1} 
\geq (\pi r_0^2 c \overline{n}_{\rm H})^{-1}  \simeq  
3\times 10^7 \, \rm yr$, for the mean density $\overline{n}_{\rm H}
\simeq 0.15 \,\rm cm^{-3}$, 
is significantly larger than both 
the Coulomb loss time and the escape times involved.   

For both CR protons and electrons  
injected into interstellar medium we assume a stationary source 
function (per unit volume)  in a `standard' power-law form with an 
exponential cutoff at some energy $E_{\rm 0}$:
\begin{equation}
q(E) \propto E^{-\Gamma_{0}} \exp(-E/E_{0})\;.
\end{equation}  

The flux of diffuse radiation with energy $E_{\gamma}$
in a given direction is defined by the unit volume
emissivity $q_{\gamma}({\bf r},E_{\gamma})$  integrated along the line of 
sight:
\begin{equation}
J(E_{\gamma})=\int \frac{q_\gamma({\bf r},E_{\gamma})}{4\pi} {\rm d} l =
\frac{\overline{q}_{\gamma}(E_\gamma)\, l_{\rm d}}{4\pi}\; , 
\end{equation}
where $\overline{q}_{\gamma}(E_\gamma)$ is the mean emissivity, and 
$l_{\rm d}$ is the 
characteristic line-of-sight depth of the emission region. 

It is convenient to describe the flux of \grs  produced at interactions 
of CRs with 
the ISM gas by  the emissivity per 1 H-atom (see e.g. Bloemen 1989). 
Then the observed intensity of \grs  linearly  depends on the column density 
$N_{\rm H}$   along the line of sight. 
It is worth  notice  that the estimate of the  mean hydrogen
density  
$n_{\rm H}^{\prime} = N_{\rm H}/l_{\rm d}$, which defines  
the mean {\it emissivities} along the line of sight in
direction close to the galactic plane,
may be somewhat higher (by a factor about 2) than the mean gas density 
$\overline{n}_{\rm H}$ which should be used for the calculation of   
the mean (at $z\leq 1\,\rm kpc$) energy distribution of
particles in Eq.(9).   
Such an `enhancement' of $\overline{n}_{\rm H}^{\prime}$
should be allowed, and could  be understood if one takes into account
that at low galactic latitudes the radiation fluxes due to CR
interactions with the gas are 
not equally contributed by the entire $z\leq 1\,\rm kpc$ region, but
effectively only by a fraction of this region close to the galactic
plane, with a thickness of a few 100 pc, where
the spatial concentrations  of {\it both} the relativistic particles and
(especially) of the ISM gas are higher
than their respective mean values averaged over $z\leq 1\,\rm kpc$. 
For calculations of the
IC $\gamma$-ray fluxes the same characteristic energy densities
of the diffuse galactic photon fields as for the Eq.(9) should be
used, because at elevations $z\sim 1\,\rm kpc$ these densities are
still approximately the same as in the Galactic plane (e.g. see Chi \&
Wolfendale 1990).

\section{Diffuse gamma radiation connected with the electronic 
component of CRs}

There are four  principal  processes of production of 
non-thermal hard X-rays  and   \grs in  the ISM by 
CR electrons:  inverse Compton (IC) scattering,
bremsstrahlung,  annihilation of  positrons, as well as  synchrotron radiation 
provided that the electrons are accelerated beyond 100 TeV. 
In this section we discuss  the first three  mechanisms; the synchrotron radiation 
of hard X-rays will be discussed in Sec.~5 in the context of the 
IC radiation of highest energy electrons.

\subsection{IC gamma rays}

The calculations of the diffuse IC $\gamma$-rays require knowledge of
the low-frequency target photons and of the flux of electrons in ISM. 

The  photon fields which are important for production  
of IC $\gamma$-rays in the ISM are 2.7\,K cosmic microwave background
radiation (MBR), and the diffuse galactic radiation 
contributed by the starlight and dust photons with peak intensities 
around $1 \, \mu \rm m$ and $100 \, \mu \rm m$, respectively.
While the  density of 2.7\,K MBR is universal, with 
$w_{\rm MBR}\approx 0.25 \, \rm eV/cm^3$,
the densities of diffuse galactic radiation  fields vary from site to site, 
and actually are model dependent.
Detailed calculations of Chi \& Wolfendale (1991) show
that the starlight energy density increases from 
the local value $w_{\rm NIR} \simeq 0.5 \, \rm eV/cm^3$ 
(Mathis et al. 1983) up to 
$\simeq 2.5 \,\rm eV/cm^3$ in the central 1\,kpc region of the
inner Galaxy. For  calculations below  we use the mean value 
$w_{\rm NIR} \simeq 1.5 \, \rm eV/cm^3$. The energy density 
of the FIR produced by dust in the galactic plane is more uncertain, 
and is typically estimated from 
$w_{\rm FIR} \simeq 0.05-0.1 \, \rm eV/cm^3$  (e.g. Mathis et al. 1983)
to $w_{\rm FIR} \simeq 0.2-0.3 \, \rm eV/cm^3$ (Chi \& Wolfendale 1991). 
For calculations of the IC \gr fluxes below we adopt  
$w_{\rm FIR} \simeq 0.2 \, \rm eV/cm^3$.
Fortunately, large uncertainties in $w_{\rm FIR}$ appear  
not crucial because at all $\gamma$-ray energies the contribution
from IC upscattering of 2.7 K target photons significantly exceeds
the IC fluxes produced on FIR  (see below).  

Another source of uncertainties in calculations of the fluxes of diffuse  
$\gamma$-rays is the lack of independent  
information about the flux and the spectrum of galactic electrons above several
GeV. While radio measurements allow definite conclusions about the average
electron flux
below a few GeV,  at higher energies the electron fluxes in the Galaxy  are 
in principle model dependent.  
The standard interpretation of the energy spectrum of CRs usually 
assumes a uniform and continuous 
distribution of sources in the Galaxy both in space and time. 
Whereas for the nucleonic component 
of CRs this approximation  can be
considered as a reasonable working hypothesis,  
the validity  of this assumption for the electrons 
is questionable at least for  
the high energy 
part of the measured spectrum which extends up to 2 TeV (Taira et al. 1993). 
Because of severe radiative losses, the sources of these electrons could 
not be located 
well beyond a few 100\,pc (Nishimura et al. 1980, Aharonian et al. 1995), 
and therefore the measured electron 
spectrum might not be applicable for calculations of   $\gamma$-radiation 
from the distant parts of the galactic disk.
%
\begin{figure}[htbp]
 \resizebox{8.5cm}{!}{\includegraphics{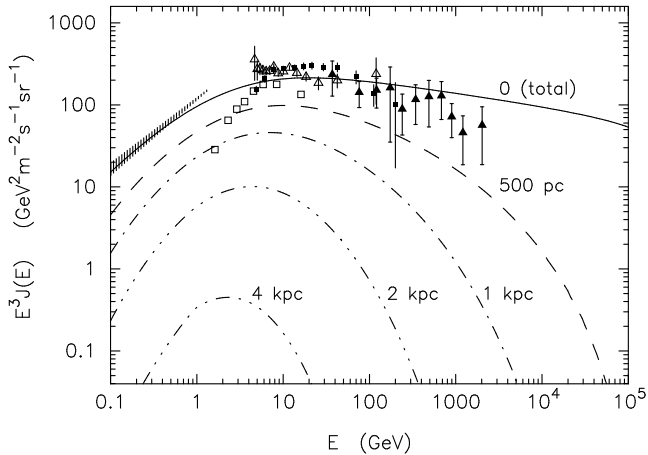}}
\caption{The fluxes of CR electrons near the Sun from sources continuously 
and uniformly distributed in the galactic disk calculated in the framework
of the diffusive propagation model for a diffusion
coefficient with $D(10\,\rm GeV)= 10^{28}\,\rm cm/s$ and a power
law index $\delta_1=0.6$, and a power-law injection spectrum of the 
electrons with $\Gamma_{\rm e,0}= 2.4$. 
The total flux (solid curve, $r_0=0\,\rm pc$) is decomposed in order 
to show the contributions
from the sources located at distances $r\geq r_0$ for
different $r_0$ indicated near
the curves. The hatched region corresponds to the estimate of the mean
flux of low energy electrons  derived from radio data in the  direction of 
galactic poles by Webber et al. (1980). The data points 
shown correspond to the local fluxes of CR electrons
measured by different groups (for details see Atoyan et al. 1995)}
\end{figure}

In Fig. 1 the energy spectrum of CR electrons calculated 
assuming a uniform and continuous 
distribution of the sources in the galactic disk (solid curve) 
is decomposed to show the contributions from sources located 
at distances $r \geq r_0$ for  different $r_0$ . It is seen that even 
at energies  $\sim 10\,\rm GeV$ the total flux of the 
observed electrons is dominated by particles accelerated and injected into
ISM at distances $r\leq 1\,\rm kpc$ from
the Sun. At TeV energies the sources beyond 500 pc 
contribute only $\sim 10 \%$ of the total electron flux.
Since  for these relatively small spatial scales the assumption of continuous
distribution (both in space and in time) of CR sources may not be well 
justified, the spectrum and the flux of high energy electrons 
at TeV and higher energies  may show significant variations in 
different sites of the galactic disk  (Atoyan et al. 1995).  
In particular, one could expect a significant 
enhancement of the electron fluxes in the central region of the Galaxy
due to presumably higher concentration of cosmic ray sources there. 
Therefore we  may allow deviations of the predicted electron distribution 
in the inner Galaxy from the observed fluxes,  perhaps 
except for the region below few GeV where the radio observations,  provide 
information  about the average spectrum of galactic electrons along the line of sight 
(see also Porter \& Protheroe 1997, Pohl \& Esposito 1998, Strong et al. 1998).
It should be noted, however, that  because of 
a significant absorption of radio fluxes in the interstellar medium, a
{\it direct} information about the spectra of CR electrons in the inner
Galaxy is not actually available. The fluxes of radio electrons
in the Galaxy are generally deduced using the  observations from the 
directions of the galactic poles or the anticenter (e.g. see Webber et al. 1980)
under an assumption of a homogeneous distribution of CRs in the galactic disk.   

In Fig.2 we show the average spectrum of electrons  in the 
inner part of the Galaxy calculated in the framework of the 
model described in section 2, assuming a power-law index for the
electron injection spectrum $\Gamma_{\rm e,0}=2.15$, and normalizing 
the energy density $w_{\rm e}=\int E\,n(E)\,{\rm d}E$ to 
$0.05\,\rm eV/cm^3$.       
For the parameters used in calculations, this normalization requires
an acceleration rate of electrons  

\begin{equation}
L_{\rm e}\simeq 1.6 \times 10^{37} \, \rm erg/kpc^3\, s \, . 
\end{equation}

This implies that for the inner part of the Galactic disk, with   
$R \leq	 8.5\,\rm kpc$ and a half thickness $ 1\,\rm kpc$, the
overall acceleration power should be about $7\times 10^{39} \,\rm
erg/s$. In the energy region   between 100 MeV and 1 GeV the energy 
losses of electrons are dominated by the adiabatic losses  and the 
bremsstrahlung,  with ${\rm d} E/{\rm d} t \propto E$,   
which do not change the original
(acceleration) spectrum of electrons. Indeed,     
it is seen from Fig.~2 that in this energy region, which is responsible 
for synchrotron radio emission, the electron spectrum  
remains rather close to the injection spectrum. 
Therefore the spectral index of the observed synchrotron
radio emission  contains almost model-independent information about the
injection spectrum of electrons.    
At energies below 100 MeV the electron
spectrum suffers 
significant deformation (flattening) because of  ionization losses, and
at energies above 1 GeV the energy distribution of the electrons
steepens because of  a combination of escape and  
radiative (synchrotron and inverse Compton) losses. 
%
\begin{figure}[htbp]
 \resizebox{8.5cm}{!}{\includegraphics{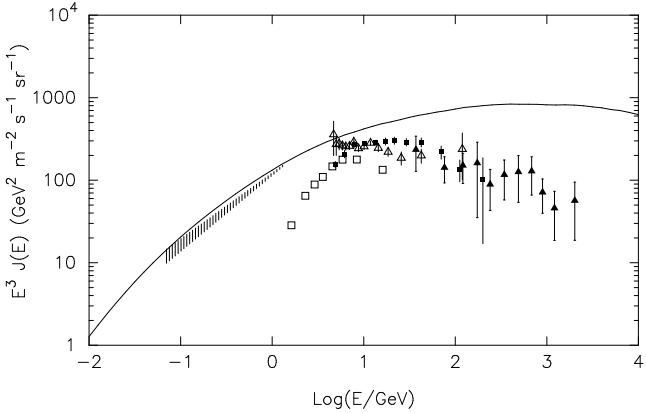}}
\caption{The mean flux of electrons in the central region of the
Galaxy calculated assuming a stationary injection spectrum of electrons 
with $\Gamma_{\rm e,0}=2.15$ and $E_{0}= 100\,\rm TeV$, and the following model
parameters for the escape time in Eq.(7): 
$\tau_{\rm conv}= 2\times 10^7\,\rm yr$,
$\tau_{10}= 10^7\,\rm yr$, $\delta =0.6$, $\tau_{\rm
min}=3\times 10^3\,\rm yr$.
For parameters of the ISM we have assumed: $B=6\,\rm \mu G$,  
$\overline{n}_{\rm H}=0.15 \,\rm H-atom/cm^{3}$ (see text), and  
$w_{\rm FIR}=0.2\,\rm eV/cm^3$,   $w_{\rm NIR}=1.5\,\rm eV/cm^3$. 
The injection rate of electrons is normalized so that the
energy density in the resulting spectrum of CR electrons is 
$w_{\rm e}=0.05 \,\rm eV/cm^3$. The hatched region is the same as in Fig.~1}
\end{figure}

\begin{figure}[htbp] 
 \resizebox{8.7cm}{!}{\includegraphics{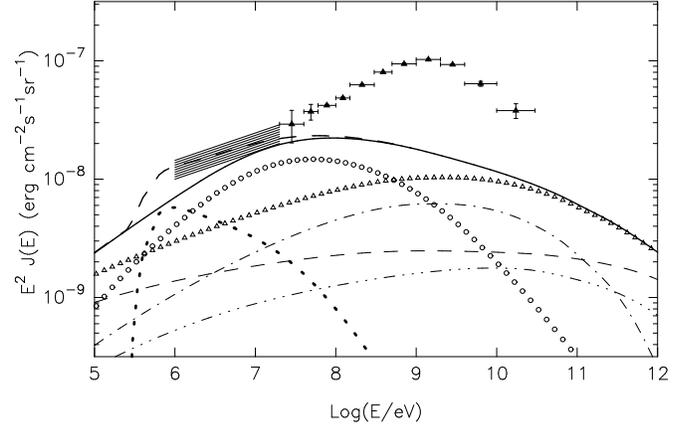}} 
\caption{The flux of diffuse $\gamma$-rays produced by the CR electrons  
due to different radiation  processes in the inner  
Galaxy. For calculations we assume the characteristic line-of-sight
depth of the emission region $l_{\rm d}=15\,\rm kpc$ and the gas column
density
$N_{\rm H}= 2 \times 10^{22}\,\rm cm^{-3}$. The open dots
show the bremsstrahlung flux, and the open triangles show the overall flux of
the IC radiation due to different target photons: 
2.7\,K MBR (thin dashed line), diffuse NIR/optical radiation assuming 
$w_{\rm NIR}=1.5\,\rm eV/cm^3$ (dot--dashed line), diffuse FIR radiation
with $w_{\rm FIR}=0.2 \,\rm eV/cm^3$ (3-dot--dashed line). 
The heavy dotted line shows the flux of $\gamma$-rays 
due to annihilation of relativistic positrons in flight 
in the case of a high charge composition $C_{+}=e^{+}/(e^{+}+e^{-})=0.5$
for electrons with energies $1-100$ MeV.   
The sum of the  bremsstrahlung and IC fluxes is shown by solid line.
The heavy dashed line corresponds to the overall \gr flux including 
also the  annihilation radiation.
The data points show the mean flux of diffuse high energy 
$\gamma$-rays observed by EGRET (Hunter et al. 1997b), and the hatched region
shows the range of average diffuse $\gamma$-ray fluxes detected by COMPTEL 
(Strong et al. 1997; Hunter
et al. 1997a ) from the direction of the inner Galaxy
at low galactic latitudes.}      
\end{figure} 

The calculated fluxes of diffuse $\gamma$-rays produced by electrons 
are shown in Fig.~3.  The spectrum of IC $\gamma$-rays below the 
highest energy  observed by EGRET, $E\leq 30\,\rm  GeV$, 
is not very sensitive to the exact value of the  
cutoff energy $E_0$  in the injection spectrum of electrons,  
provided that  $E_0$  exceeds 10 TeV.
For the energy density of the diffuse interstellar  NIR/optical radiation   
we have assumed  $w_{\rm NIR} = \rm 1.5 \, eV/cm^3$.
For this value of  $w_{\rm NIR}$  the IC radiation 
component produced on the galactic starlight photons 
(dot-dashed line) somewhat exceeds 
in the energy region 10 MeV - 30 GeV  the IC flux produced on 
 2.7\,K MBR (dashed line). The 
`FIR' component of IC radiation (3-dot--dashed line) calculated for  
$w_{\rm FIR} = \rm 0.2 \, eV/cm^3$  at any $\gamma$-ray energy 
contributes less than  $25 \%$  of  the total IC flux.

In its turn, the overall  IC $\gamma$-ray flux can account,
for the chosen infrared photon field  densities,  only 
for $\leq 20 \,  \%$  of the observed $\gamma$-ray fluxes both at MeV 
(``COMPTEL'') and GeV  (``EGRET'') energies (see Fig.~3). For the same 
average electron fluxes shown in Fig.2, the fluxes of IC radiation could
be increased assuming formally a larger  depth $l_{\rm d}$ of the emission 
region. However,  the  value of the mean $l_{\rm d}= 15 \,\rm kpc$
assumed in Fig.3 is already large,
and hardly it could be significantly increased further.
Another way to increase the flux of IC $\gamma$-rays
 is possible if we assume that the  energy density 
of the electrons in the inner Galaxy is significantly larger than 
$w_{\rm e}=0.05\,\rm eV/cm^3$.  
However, for a fixed gas column density $N_{\rm H}$
this would automatically increase also
the flux of the bremsstrahlung $\gamma$-rays, resulting in an
overproduction of diffuse radiation in the  $10-30 \,\rm MeV$ region 
(see below).  

The possibilities to increase the flux of IC $\gamma$-rays to a level
significantly higher than in Fig.3  are essentially  limited also by the
radio observations.
The density of the Galactic electrons in the energy range 
70\,MeV to  1.2\,GeV  shown in Fig.~2 is derived by Webber et al. (1980)
from radio observations at low frequencies in the galactic pole directions 
assuming the average magnetic field $B \sim 6\,\rm \mu G$. 
Thus, below few GeV the spectral index 
of electrons is
well fixed, $\Gamma_{\rm e}=2 \,  \alpha_{\rm r} -1=2.14\pm 0.06$
(for the photon spectral index of the 
observed radio emission $\alpha_{\rm r}=1.57 \pm 0.03$),  
but their absolute flux depends on the magnetic field.

The mean energy of the 
IC $\gamma$-rays produced by an electron with energy $E_{\rm e}$  on 
target photons with an energy $\epsilon_0$ is 
$E_{\rm IC}=(4/3)\, (E_{\rm e}/m_{\rm e}c^2)^2\,\epsilon_0$ . 
Therefore IC radiation of 1\,GeV `radio' electrons on the IR/optical 
photons with $\epsilon_0\sim 1-2\;\rm eV $ corresponds to energies  
$E\sim  5-10 \; \rm MeV$. The expected energy 
flux,  $F(E)=E^2 J(E)$,  of these
IC $\gamma$-rays can be estimated analytically: 
\begin{eqnarray}
F_{\rm IC}^{\rm (NIR)}(E)& \simeq & 1.5\times 10^{-9}\; 
\frac{w_{\rm e}}{0.1\,\rm eV/cm^3}\;
\frac{w_{\rm NIR}}{1\,\rm eV/cm^3} \;
\frac{l_{\rm d}}{10\,\rm kpc} \nonumber \\
& & \times  
 \left(\frac{\epsilon_0}{1\,\rm eV}
\right)^{\frac{\Gamma_{\rm e}-3}{2}}
\left( \frac{E}{ 5\,\rm MeV}
\right)^{\frac{3- \Gamma_{\rm e}}{2}} 
\frac{\rm erg}{\rm cm^2\, s\, sr}\, 
\end{eqnarray}
for $\Gamma_{\rm e} \simeq 2.15$. 
The comparison of Eq.(14) with the results of numerical calculations  
(the dot-dashed line in Fig.3) shows a reasonable accuracy of this convenient 
analytical expression. Normalizing the flux of GeV electrons to the
radio flux , 
$F_{\nu}= \nu S_\nu \simeq 10^{-10} \, \rm erg/cm^2 \, s \, ster$ 
at $\nu=10\,\rm MHz$, 
we find a direct relation between the IC
 $\gamma$-ray fluxes produced on NIR/optical photons by GeV electrons and
 the {\it un-absorbed} radio fluxes to be expected from the inner Galaxy:  
\begin{eqnarray}
F_{\rm IC}^{\rm (NIR)}(E)& \simeq & 2.6 \;
\frac{w_{\rm NIR}}{1\,\rm eV/cm^3} 
\left(\frac{B}{6\,\rm \mu G}\right)^{-\frac{1+\Gamma_{\rm e}}
{2}} \times \nonumber \\
& & 
 \left(\frac{\epsilon_0}{1\,\rm eV}
\right)^{\frac{\Gamma_{\rm e}-3}{2}}
\left( \frac{E}{ 5\,\rm MeV}
\right)^{\frac{3 - \Gamma_{\rm e}}{2}}
 F_{10\,\rm MHz} \, .
\end{eqnarray}  
For a given  radio flux in the galactic plane,
a decrease of the magnetic field by a factor of two
would lead to  an increase
of the electron flux shown in Fig.~2 by a factor of 
$2^{\alpha_{\rm r}} \sim 3$, 
 and correspondingly to the  
increase of the IC $\gamma$-ray fluxes by the same factor.   

Unfortunately, at low frequencies the absorption of radio fluxes from 
the inner Galaxy at low galactic latitudes is very significant, so the
diffuse radio flux $F_{10 \,\rm MHz}$ is uncertain.  Eq.(15) 
predicts that this analytical estimate of the IC flux (for $E \leq 
10 \,\rm MeV$)  would be in agreement with the 
results of numerical calculations shown by the dot-dashed line in Fig.~3
for the un-absorbed radio flux from the inner Galaxy 
$F_{10 \,\rm MHz} \sim 4\times 10^{-10}\,\rm erg / cm^2 \, s \, ster $,
i.e. by a factor 5  larger than the Galactic radio flux 
observed in the polar directions, $\approx 8 \times 10^{-11}
\,\rm erg / cm^2 \, s \, ster $ (Webber et al.1980).
This implies a size of the halo of the Galactic CRs   
extending up to heights $\sim 3\,\rm kpc$ (to be compared with 
$l_{\rm d} \sim 15 \,\rm kpc $ used in Fig.~3), which is
in agreement with the relevant theoretical predictions in the framework 
of the diffusion models (e.g. Bloemen et al. 1993).

Fig. ~3 and Eq.\,(14) show that any attempt to explain 
the observed $\gamma$-ray  fluxes at $E \sim 1-10 \, \rm MeV$, 
$F_{\rm obs}\geq 10^{-8} 
\,\rm erg\, cm^{-2}\, s^{-1}\, sr^{-1}$ by IC radiation either would
require an energy density of the NIR/optical radiation at a level of 
$\geq 5 \, \rm eV/cm^3$, which is much larger than it is generally accepted for the  
density  of starlight  photons in the ISM, 
or would require a  very high flux of the radio electrons in the inner Galaxy.
The latter assumption  would imply, however, very high 
radio fluxes, exceeding by more than one order of magnitude the flux
detected from  the direction of the Galactic poles, unless we assume 
a (unrealistically)   low magnetic field, 
$B \sim 1 \mu \rm G$ or so, in the Galactic plane. 
Moreover,  independently  of the strength of the interstellar magnetic field,
such an assumption of high flux of $E\leq 1\,\rm GeV$ electrons leads to 
a simultaneous increase and  overproduction of  the bremsstrahlung flux as 
well, which would then exceed the $\gamma$-ray 
flux observed between 1 and 100 MeV.     

\subsection{Electron bremsstrahlung}
 
Since the bremsstrahlung $\gamma$-rays below 1 GeV are produced by the 
same electrons which are responsible also for the galactic synchrotron 
radio emission, the differential flux $J(E)$ of this radiation
in the region from $30 \,\rm MeV$ to $\sim 1\,\rm GeV$ should 
have a characteristic power-law slope with an 
index coinciding with the spectral 
index of the  radio electrons 
$\Gamma_{\rm e}\simeq 2.1-2.2$. The  results of 
numerical calculations are  shown  in  Fig.~3  by open dots. For a
power-law spectrum of electrons the energy flux of  bremsstrahlung
$\gamma$-rays can be calculated analytically:   
\begin{eqnarray}
F_{\rm brem}(E)& \simeq & 1.3 \times 10^{-8}\; 
\frac{w_{\rm e}}{0.1\,\rm eV/cm^3}\;
\frac{N_{\rm H}}{10^{22}\,\rm cm^{-2}} \,
\times \nonumber \\
& & 
\left( \frac{E}{ 100 \,\rm MeV}
\right)^{2-\Gamma_{\rm e}} \;
\frac{\rm erg}{\rm cm^2\, s\; sr}\; ,
\end{eqnarray}
In calculations we have assumed a 
standard composition of the interstellar gas 
 ($\simeq 90\,\%$ of the molecular and atomic hydrogen,
and $\simeq 10\,\%$ of helium).

It is worthwhile to compare the bremsstrahlung flux  
in the energy region $E\sim 30\,\rm MeV$ with the `NIR' component
of IC flux in the region $E\sim 10\,\rm MeV$ since both
components are due to the  radiation of the same  radio electrons, 
(although contributed by two different,  low-energy and  high-energy, 
parts of the power-law distribution of radio electrons,  respectively).  
Assuming for the mean photon energy of NIR $\epsilon_0 \simeq 1\,\rm eV$, 
and $\Gamma_{\rm e}=2.15$ for the radio electrons,
from Eqs.~(14) and (16) we find:
\begin{eqnarray}
\frac{F_{\rm brem}(30\,\rm MeV)}{F_{\rm IC}^{\rm (NIR)}(10\,\rm MeV)}
 & \simeq & 7.6 \, \frac{N_{\rm H}}{10^{22}\,\rm cm^{-2}}
\left( \frac{ l_{\rm d}}{10\,\rm kpc} \right)^{-1} \times \nonumber \\ 
&  & 
\left( \frac{w_{\rm NIR}}{1\,\rm eV/cm^3}\right)^{-1}\, \; . 
\end{eqnarray}

Comparison of Eq.(15) with the results of numerical calculations in 
Fig.3 shows a good accuracy of this analytical estimate.  

In the region $E\leq 30\,\rm MeV$ the bremsstrahlung flux is due to 
electrons with $E_{\rm e} < 70\,\rm MeV$ (i.e. outside the domain 
of radio emitting electrons) where ionization losses 
result in a significant flattening
of the electron spectra $n(E_{\rm e})$. This results in a drop of
$F_{\rm brem}$ at 10 MeV by a factor of 1.5 compared with 
$F_{\rm brem}(30\,\rm MeV)$. 
For $w_{\rm NIR} = 1.5 \, \rm eV/cm^3 $ assumed in Fig.~3 
the overall IC flux at 10 MeV is comparably contributed by both
NIR and MBR target photons. In the case of a 
higher density of the diffuse NIR field in the inner Galaxy the overall IC
radiation at those $\gamma$-ray energies will be dominated by the IC 
upscattering of the starlight photons. Taking all these effects into account, 
one can conclude from Eq.(17) that at energies
$E\sim 10\,\rm MeV$ the bremsstrahlung
should dominate the overall diffuse emission observed in the
direction of the Galactic plane,
unless one assumes a very high energy density of NIR, 
$w_{\rm NIR} \geq 5 \, \rm eV/cm^3$ 
as adopted by Strong et al. (1998).  

Independently of the density of the NIR 
in the inner Galaxy, a conclusion that the contribution of
the brems-
strahlung to the overall flux of the galactic diffuse 
$\gamma$-ray background is large,  can be derived from the comparison of 
the fluxes produced by the same $70\,{\rm MeV} \leq 
 E_{\rm e}\leq 1\,\rm GeV$ electrons 
in the radio and $\gamma$-ray regions. At the photon energy $30\,\rm MeV$
this results in 
\begin{eqnarray}
F_{\rm brem}(30\,{\rm MeV}) & \simeq  & 
28 \, \frac{N_{\rm H}}{10^{22}\,\rm cm^{-2}}
\left( \frac{ l_{\rm d}}{10\,\rm kpc} \right)^{-1} \times \nonumber \\  
& &  \left(\frac{B}{6\,\rm \mu G}\right)^{-1.57}
 F_{10\,\rm MHz} \; ,
\end{eqnarray}
where $F_{10\,\rm MHz}$ is an  un-absorbed flux produced in the galactic 
plane. Since the latter cannot be less (and, presumably, is even 
several times higher) than the flux 
$10^{-10} $ $ \rm erg \, \rm cm^{-2}\, s^{-1} \,sr^{-1}$
detected from the direction of the Galactic poles (Webber et al. 1980), 
the contribution of the  bremsstrahlung to the overall diffuse flux of 
$E\sim 10\,\rm MeV$ $\gamma$-rays should be significant,  unless 
one would assume either a very high   magnetic field  
($B\gg 10\,\rm \mu G$) or a very low  gas column density 
($N_{\rm H} \ll  10^{22} \, \rm cm^{-2}$)   in the direction of the 
inner Galaxy.

\subsection{Annihilation of CR positrons in flight}

Since a significant fraction of the CR electron component could be
in the form of positrons, a non-negligible contribution to the
diffuse $\gamma$-radiation below 10 MeV could be due to  
annihilation of mildly relativistic positrons (Aharonian \& Atoyan 1981a,
Aharonian et al. 1983). The differential spectrum of the $\gamma$-rays 
produced at the annihilation of a fast positron with a Lorentz-factor 
$\gamma_{+}=E_{+}/m_{\rm e}c^2$ on the ambient electrons with density
$n_{\rm e}$  
is described by a simple analytical expression (Aharonian \& Atoyan 1981b) 
\begin{eqnarray}
q_{\rm ann}(\varepsilon)& = & \frac{\pi r_{\rm e}^2 c n_{\rm e}}{\gamma_{+}\,
p_{+}}
\left[ 
\left( \frac{\varepsilon}{\gamma_{+}+1 -\varepsilon}
+\frac{\gamma_{+}+1 -\varepsilon}{\varepsilon} \right)
 + \right. \\
& & \left.
2 \left(\frac{1}{\varepsilon} +\frac{1}{\gamma_{+}+1-\varepsilon}\right) - 
\left(\frac{1}{\varepsilon} +\frac{1}{\gamma_{+}+1-\varepsilon}\right)^2
\right] \nonumber
\end{eqnarray}
where $p_{+}=\sqrt{\gamma_{+}^2-1}$ is the dimensionless momentum
of the annihilating positron\footnote{Note that in Aharonian \& Atoyan (1981b) 
there is a misprint in Eq.(5), namely the momentum $p_+$  in the denominator
is missing.}, and the photon energy 
$\varepsilon=E/m_{\rm e}c^2$ varies in the limits
\begin{equation}
\gamma_{+}+1-p_{+}\leq 2 \varepsilon\leq 
\gamma_{+}+1+p_{+}\; .
\end{equation}

For the power-law spectrum of the positrons 
$N_{+}\propto \gamma_{+}^{-\Gamma_{\rm e}}$, 
the spectrum of annihilation radiation at high energies  
$E \gg m_{\rm e}c^2$
has a power-law form
\begin{equation}
J_{\rm ann}(E)\propto E^{-(\Gamma_{\rm e}+1)}
\,[\ln(2E/m_{\rm e}c^2)-1]\;. 
\end{equation}
Thus, the spectrum of annihilation radiation is steeper
than the spectrum of electrons, in contrast to  
the spectrum of bremsstrahlung
radiation which in the high energy limit repeats the spectrum of parent electrons.
At lower energies the spectrum has a more complicated form with a maximum 
around 1 MeV. 
The ratio of the fluxes $J_{\rm ann}/J_{\rm brem}$ does not depend on the 
ambient gas density,
and for  a given ratio $C_{+}=e^{+}/(e^{+}-e^{-})$
depends only on the spectrum of electrons,
being higher for steep electron spectra.

The flux of annihilation radiation calculated for the spectrum 
of electrons shown in Fig.~2,  and assuming  $50 \%$ content  of positrons, 
is presented in Fig.~3.  
It is seen that under such an assumption  the contribution of the 
annihilation radiation at 1 MeV exceeds the fluxes  produced by all
other radiation processes, including the bremsstrahlung.  Therefore  we
conclude that depending on the (unknown) content of low-energy ($\leq
100 \,\rm MeV$) positrons in CRs, 
this process may result in a significant enhancement of
the diffuse radiation at MeV energies. 
Note that at energies $E_{\rm e}< 1 \,\rm GeV$ the
fraction of positrons in the local (directly measured)  
component of CR  electrons gradually increases, reaching  
$C_{+} \geq 0.3 $ (although with large uncertainties) at 
$E_{\rm e} \sim 100 \,\rm MeV$  (Fanslow et al. 1969). 
A detailed discussion of different possibilities which may provide 
enhanced positron flux at low energies is  out of the scope 
of this paper. 
We note only that pulsars could be potential suppliers of
low-energy $\rm (e^+,e^-)$ pairs into the electronic component of 
CRs (see e.g. Harding \& Ramaty 1987). Also, we may speculate  that 
a moderate acceleration of beta-decay positrons produced at early stages 
in SNRs would result in such an enhancement. Obviously 
these possibilities require  thorough examination, therefore the adopted 
in this paper  large content of positrons at  low energies should be considered
as a {\em working hypothesis} which helps to explain the MeV excess 
in the galactic diffuse background radiation.

The assumption that the process of annihilation of supra-thermal 
positrons in flight
might significantly  contribute to the diffuse low-energy  
$\gamma$-radiation of the inner Galaxy in principle would imply also a 
high flux of 0.511\,MeV  annihilation line radiation,
and  a rather broad continuum
emission at $E\leq 0.511 \,\rm MeV$ due to annihilation of the
thermalized  positrons through the positronium channel.  
 OSSE measurements (Purcell et al. 1993) have  shown 
that the flux of the diffuse annihilation radiation detected from the inner 
Galaxy is dominated  ($\simeq 97\, \%$, Kinzer et al. 1996) by the
positronium annihilation, which consists of two components - a 
narrow 0.511 annihilation line  and  the
$\gamma$-ray continuum below 0.5 MeV.  
At energies $E\sim (0.2-0.5) \,\rm MeV$
the diffuse $\gamma$-ray emission of the
inner Galaxy is  contributed mainly by the fluxes of these two components
of the annihilation radiation  (not shown in Fig.~3;  but see
e.g. Hunter et al. 1997b). 
The total flux of these photons  is at the level    
$\simeq 10^{-3} \,\rm cm^{-2} s^{-1}$ (see Kinzer et al. 1996), 
which is equivalent to $\simeq 2.4 \times 10^{-2}\,\rm cm^{-2} \,s^{-1} 
ster^{-1}$ for the field of view $3.8^{\circ} \times  11.4^{\circ}$
of the OSSE instrument. 

Our earlier calculations (Aharonian \& Atoyan 1981a) show  that in the
approximation of an infinite interstellar medium,  about $ 20\,\%$ 
of relativistic positrons  would annihilate in flight on the
thermal electrons of the same gas medium where they cool due to 
(predominantly) Coulomb and bremsstrahlung energy losses.
Thus, in this case the total photon flux from the  annihilation of
the positrons after their thermalization in the ambient ISM would be  
by a factor of  4  larger than the total photon flux due the 
annihilation of relativistic positrons. The integrated flux of the annihilation 
radiation by relativistic electrons shown in Fig.~3 (dotted curve),   
is $\simeq 6.7 \times 10^{-3} \,\rm cm^{-2}\,s^{-1} \, ster^{-1}$.
Therefore one could expect that the photon flux associated 
with the thermalized component of CR positrons would be as high as   
$J_{0.511} \simeq 2.7  \times 10^{-2} \,\rm cm^{-2}\,s^{-1} \, ster^{-1}$, 
provided that the  positron content in the total flux of CR  
electrons is as high as $C_{+} =0.5$, as assumed in Fig.~3. This flux is
quite comparable with the annihilation radiation flux observed.

We should note, however, that the ratio of CR positrons that annihilate
after their thermalization to the positrons annihilating while remaining
still relativistic may be in fact significantly (by a factor of few)  
lower because this estimate does not take into account  
the escape losses of particles from the thin gaseous disk of
the Galaxy. Meanwhile, the positrons, both relativistic and thermalised, 
escape (in particular, by convection) from the disk on timescales
comparable with, and even shorter than their cooling and annihilation
times, which may therefore significantly reduce (for the same
high $C_{+}$) the flux of the annihilation radiation associated  
with the thermalized CR positrons. This problem needs, however, a
separate study which is out of the scope of the present paper. 

A possible source of relativistic positrons in the energy region below 100 MeV, 
which presents a prime interest from the point of view of production of the 
continuum annihilation  
radiation (by relativistic positrons)  and the subsequent production 
of the 0.511 MeV  line and three-photon positronium continuum 
(by thermalized positrons), are 
interactions of CR protons and nuclei  with the ambient gas 
via production and decay
 of  secondary $\pi^+$-mesons.  The total $\gamma$-ray flux above 100 MeV 
from the inner Galaxy, associated with the  $\pi^0$-decay component of 
diffuse radiation, cannot significantly exceed 
$\sim 10^{-4} \, \rm ph/cm^2 \ s \ sr$ (see below), therefore it strongly 
limits the contribution  of  these positrons to the
observed flux of 0.511 MeV line, and consequently also 
to the $\gamma$-ray continuum at $E \leq 10 \, \rm MeV$.  
Obviously a more copious mechanism for production of  
positrons in the Galaxy with energy $\leq 100 \, \rm MeV$ is needed
in order to interpret the ``MeV'' excess of the 
diffuse $\gamma$-radiation of the inner Galaxy.
Ejection of relativistic electron-positron pairs from the pulsar 
magnetospheres seems an interesting possibility. Apparently, 
this question also  needs a separate detailed study, 
which cannot be done in this paper.

\vspace{2mm}

The overall flux of bremsstrahlung and IC diffuse $\gamma$-rays
is shown in Fig.~3 by solid line.  It is seen that the $\gamma$-radiation 
of the CR electrons is significantly below the measured spectrum in the 
entire energy range  from 100\,MeV to  30\,GeV.  
Due to the lack of independent information on the spectrum of high energy
electrons with $E_{\rm e} \gg 1 \, \rm GeV$ in the inner Galaxy,
the predictions of diffuse $\gamma$-radiation above several 100 MeV
could contain  significant uncertainties. Although both at 100 MeV
and 30 GeV energies the deviation of the calculated fluxes from the 
measurements (by a factor of $\approx 2)$ should not be  
overemphasized,  the gap by a factor of 
5 to 7 around 1\,GeV is  not easy to explain by a reasonable  set of model parameters. 
Moreover, the flat shape of the  overall flux produced by CR electrons 
cannot explain the GeV bump without violation of the 
fluxes observed at lower energies. Below we study the possibility of 
explanation of this bump   
by the nucleonic component of diffuse radiation connected with interactions of 
CR protons and nuclei  with the interstellar gas.

\section{Gamma rays of nucleonic origin}

 \subsection{Emissivity of $\pi^0$-decay $\gamma$-rays}

Relativistic protons and nuclei produce  \grs  in the inelastic collisions 
with ambient nucleons due to production and decay of $\pi^0$-mesons, 
$pp \rightarrow \pi^0\rightarrow 2 \gamma$. This mechanism 
has been extensively studied   by many authors 
(e.g. Stecker 1979, Dermer 1986, Berezinsky et al. 1993,
Mori 1997). Here
we present a simple formalism which allows us to calculate with high accuracy
the emissivity of $\gamma$-rays in the case of any broad energy 
distribution of CRs.

The emissivity $q_{\gamma}(E_{\gamma})$ of $\gamma$-rays due to decay
of $\pi^0$-mesons is directly defined  by their emissivity $q_{\pi}(E_{\pi})$ as
\begin{equation}
q_{\gamma}(E_{\gamma}) = 2 \int_{E_{\rm min}}^{\infty}
\frac{q_{\pi}(E_{\pi})}{\sqrt{E_{\pi}^2- m_{\pi}^2 c^4}} {\rm d}E_{\pi} \; ,
\end{equation}
where $E_{\rm min}= E_{\gamma}+ m_{\pi}^2 c^4/4 E_{\gamma}$, and 
$m_{\pi}$ is the $\pi^0$-meson rest mass. 

The emissivity of secondary particles from inelastic proton-proton interactions
can be calculated with high accuracy 
using accelerator  measurements of  the inclusive cross-sections 
$\sigma (E_{i}, E_{\rm p})$ for production of a particle $i$ 
in hadronic interactions (see e.g. Gaisser 1990).  
The emissivity of $\pi^0$-mesons calculated in the $\delta$-functional 
approximation for the cross-section $\sigma (E_{\pi}, E_{\rm p})$ reads
\begin{eqnarray}
q_{\pi}(E_{\pi})& = & c\, n_{\rm H} \int \delta (E_{\pi}-K_{\pi} E_{\rm kin}) 
\sigma_{\rm pp} (E_{\rm p}) n_{\rm p}\! (E_{\rm p}){\rm d}E_{\rm p} 
\nonumber \\ 
 & = & \frac{c\, n_{\rm H}}{K_{\pi}} 
\sigma_{\rm pp} \! \left( m_{\rm p} c^2 +\frac{E_\pi}{K_\pi}\right)\,
n_{\rm p}\! \left( m_{\rm p} c^2 +\frac{E_\pi}{K_\pi}\right) 
\end{eqnarray}
where $\sigma_{\rm pp}(E_{\rm p})$ is the total cross section of inelastic 
$pp$ collisions, and $K_\pi$ is the mean fraction of the 
kinetic energy $E_{\rm kin}= E_{\rm p}-m_{\rm p}c^2$ of the proton 
transferred to the secondary $\pi^0$-meson per collision; 
$n_{\rm p}(E_{\rm p})$ is the energy distribution of the protons.

In a broad region from GeV to TeV  energies $K_\pi\approx 0.17$
which includes also $\sim 6\%$ contribution from $\eta$-meson production
(Gaisser 1990). From the threshold at $E_{\rm kin} \simeq 0.3 \,\rm GeV$, 
 $\sigma_{\rm pp}$ rises rapidly to 
about $28-30\,\rm mb$ at energies about $E_{\rm kin}\leq 2 \,\rm GeV$.
After that $\sigma_{\rm pp}$ increases  only logarithmically.
For calculations we approximate 
\begin{equation}
\sigma_{\rm pp}(E_{\rm p})\approx 30\, [0.95 +0.06\,\ln 
(E_{\rm kin}/1\,\rm GeV)] \; \rm mb\; 
\end{equation}
for $E_{\rm kin}\geq 1\,\rm GeV$, and assume $\sigma_{\rm pp}=0$
at lower energies.  More accurate approximation of the 
cross-section  below  1 GeV  (see e.g. Dermer 1986) does not noticeably change the fluxes
of \grs even at very low energies provided that 
the broad power-law spectrum of protons extends beyond 10 GeV, and thus
the overall flux is contributed by protons with energies above few GeV. 
 
Good accuracy of this simple approach is demonstrated 
in Fig.~4 where we compare the emissivity of $\pi^0$-decay 
$\gamma$-rays  calculated on the base of Eqs.(22)-(24)  
with the results of  the recent Monte-Carlo 
calculations of Mori (1997) based on a      
detailed treatment of the cross-sections of  
secondary pion production at  nucleon-nucleon  interactions. 
The full dots in Fig.~4 correspond to calculations
of Mori for his ``median'' proton flux (Eq.~3 in Mori 1997). 
The solid line in Fig.~4  corresponds to our calculations for the 
same ``median'' proton flux. 

The dashed curve corresponds to our calculations  
but for the local CR proton flux (see e.g. Simpson 1983)  in the form 
\begin{equation}
J_{\odot}(E_{\rm p})=2.2 (E_{\rm p}/1\,{\rm GeV})^{-2.75}\; 
\rm cm^{-2}s^{-1}sr^{-1} GeV^{-1}\; ,
\end{equation}
which has been used for the detailed $\gamma$-ray emissivity
calculations 
by Dermer (1986). Here also we have a very good  agreement;  
the original spectrum from Dermer (1986) is not shown in order not 
to overload the  figure with almost coinciding curves.

\begin{figure}[htbp]
 \resizebox{7.5cm}{!}{\includegraphics{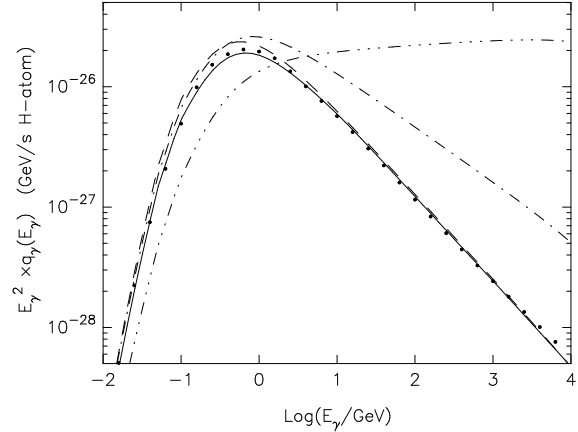}}
\caption{ The emissivities, per 1 H-atom, of $\pi^0$-decay $\gamma$-rays 
calculated using 
an approximate method given by Eqs.~(22)--(24) for the spectra of CR protons
corresponding to the `median' proton flux (solid line) of Mori (1997), and 
the flux given by Eq.~(25) (dashed line), as compared with the results of
detailed emissivity calculations by Mori (1997) shown by full dots.  
The dot-dashed and 3-dot--dashed curves correspond to the emissivities
calculated for the single power-law spectra of protons with 
indices $\Gamma_{\rm p}=2.5$ and $\Gamma_{\rm p}=2$.  
}
\end{figure}

Two other curves in Fig.~4 correspond to the emissivities calculated for
the power law proton spectra with spectral indices $\Gamma=2.5$ (dot--dashed
curve) and $\Gamma=2$ (3-dot--dashed curve), normalized to the
same energy density of CR protons $w_{\rm p}=
\int n_{\rm p}(E_{\rm p}) E_{\rm p}{\rm d}E_{\rm p}\approx 1.2 \,\rm eV/cm^3$ 
derived from the  CR proton flux given by Eq.(25). 
For CR proton spectra with 
$\Gamma \geq 2.4$ the emissivities, in terms of $E_{\gamma}^2 \,q_{\gamma}$,
reach the maximum at $E \simeq 1 \, \rm GeV$, and then at lower energies
the spectra sharply decline. Note that we do not find 
any peculiarity  in the declining part of the spectrum at energies between 100 MeV
and 1 GeV neither in our nor in the  Dermer's (1986) or Mori's (1997) 
spectra, in contrast to the apparent changes of
the sign of the second derivative in the emissivity spectra presented by 
Pohl \& Esposito (1998) and Strong et al. (1998).

\subsection{Fitting the GeV bump}

An important feature of the gamma-ray of nucleonic origin which 
allows to fill up the observed `bump' of diffuse radiation at GeV
energies
in Fig.~3 without 
violation of the fluxes observed at MeV energies 
is the profound drop of the spectrum of this  this component 
below 100 MeV. 

In  Fig.~5a we show the fluxes calculated for a single power-law 
spectra of CR protons $J(E_{\rm p}) \propto E_{\rm p}^{-\Gamma_{\rm p}}$ assuming
different values for $\Gamma_{\rm p}$. All fluxes shown correspond to
the product of the energy density of protons to column density of gas
in the inner Galaxy 
$w_{\rm p} \,  N_{\rm H}= 2.5\times 10^{22}\,\rm eV/cm^5$. Besides, 
in order to take into account a contribution due to CR nuclei, 
hereafter the fluxes  produced by CR protons are multiplied  
by a {\it constant} nuclear enhancement factor 
$\eta_{\rm A}=1.5$ which takes into account the contribution from
the nuclei both in CRs and ISM  
(Dermer 1986), although at energies above 100 GeV this factor may gradually 
increase (see Mori 1997). 

\begin{figure}[htbp]
 \resizebox{7.5cm}{!}{\includegraphics{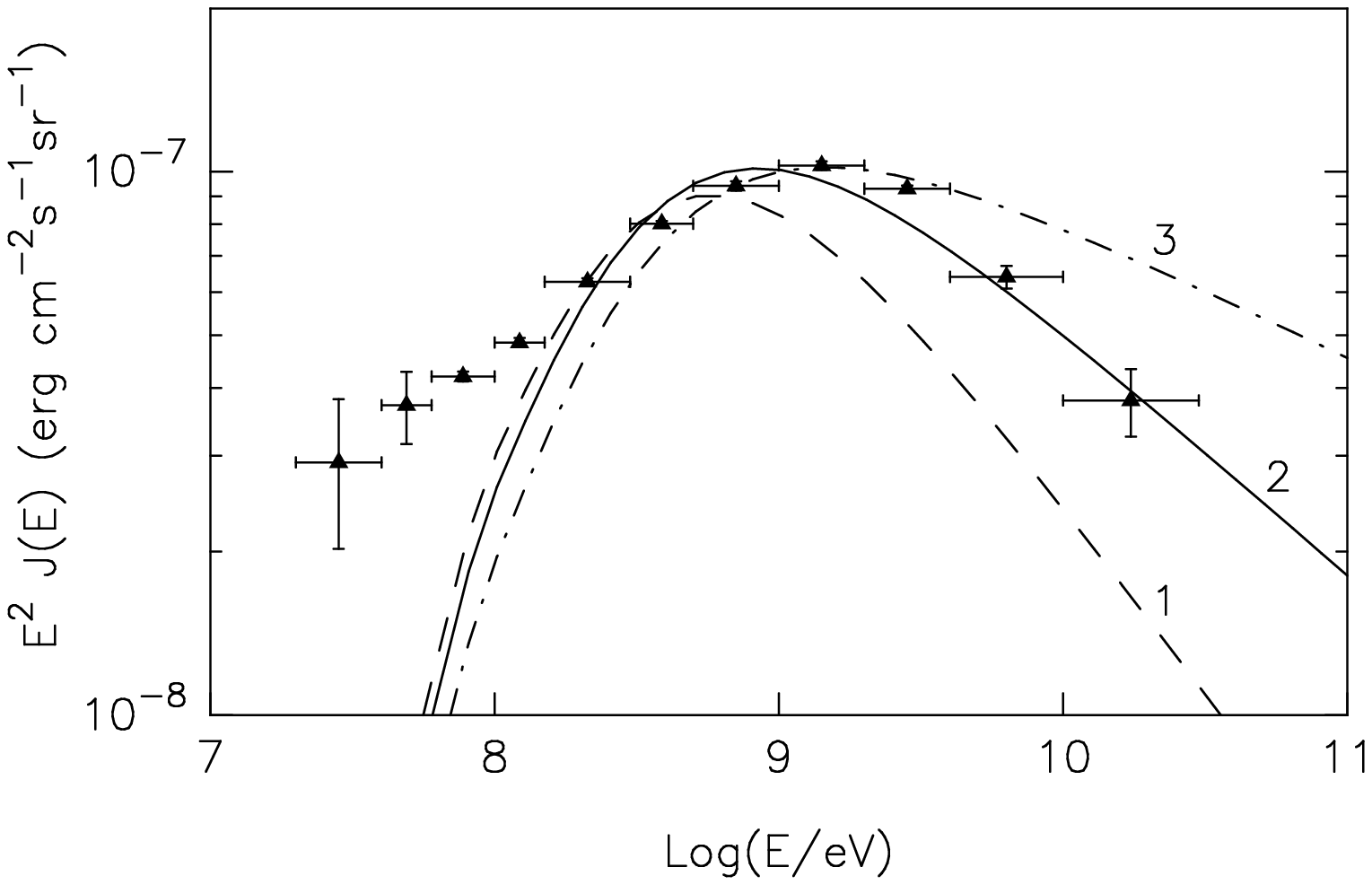}}
 \resizebox{7.5cm}{!}{\includegraphics{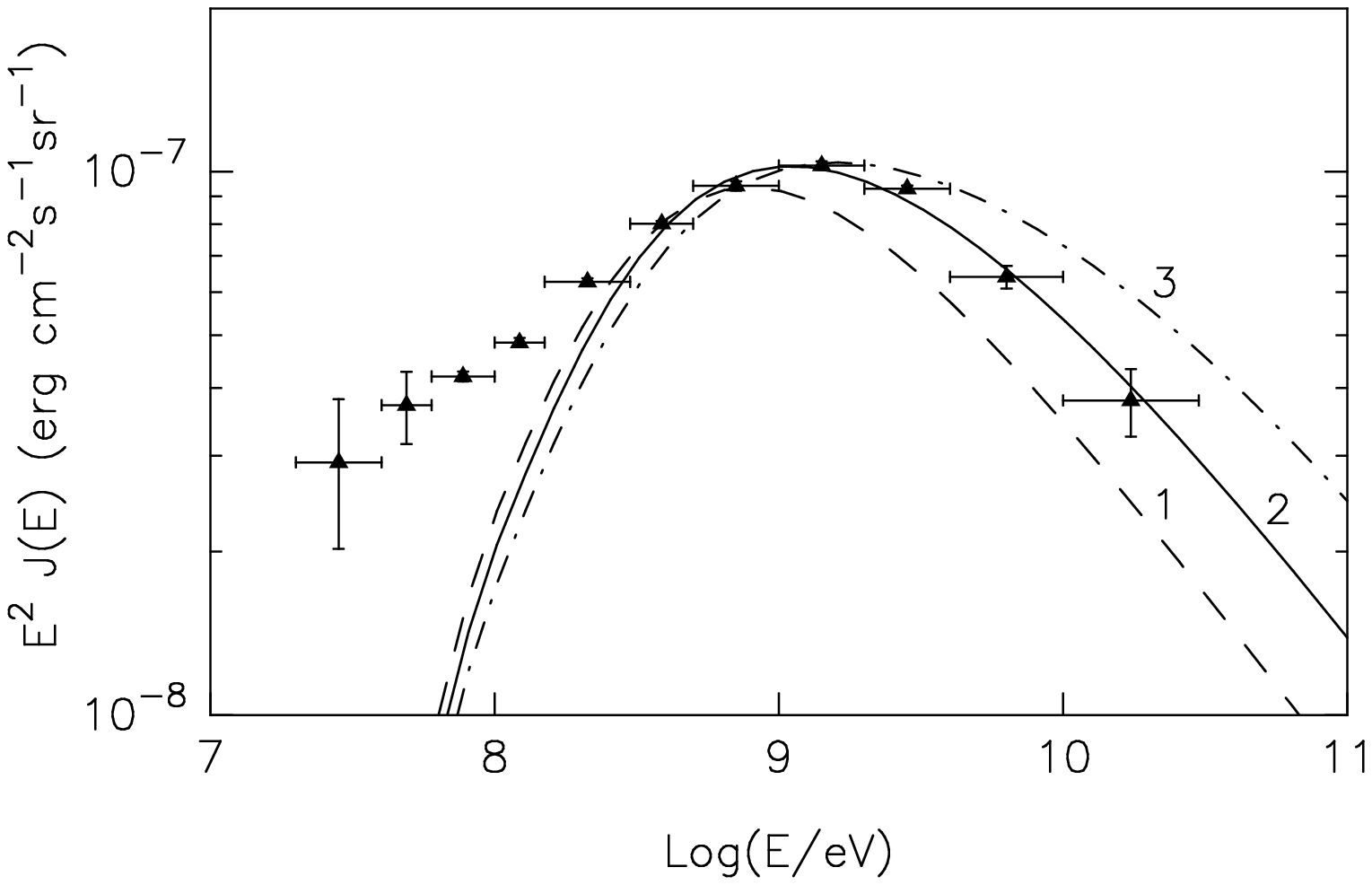}}
\caption{ The fluxes of $\pi^0$-decay $\gamma$-rays calculated for CR proton
energy distributions given in a single power-law form (Fig.~5a - top panel),
and in  the form of Eq.(26) (Fig.~5b - bottom panel). 
The fluxes in Fig.~5a are calculated 
assuming for the product  $w_{\rm p}\,N_{\rm H}=2.5\times
10^{22}\,\rm eV/cm^5$ and 3 different power-law indices 
$\Gamma_{\rm p}$: 2.75 (curve 1 -- dashed), 2.5 (2 -- solid), and 2.3 
(3 -- dot-dashed).
The fluxes in Fig.~5b are calculated  assuming
$w_{\rm p}\,N_{\rm H}=2.1\times 10^{22}\,\rm eV/cm^5$, and the same 
indices $\Gamma_{0}=2.1$ and $\delta=0.65$ in Eq.(26), but three
different
values for the characteristic energy $E_{\ast}$: 3\,GeV (1 -- dashed), 
20\,GeV (2 -- solid), and 100\,GeV (3 -- dot-dashed). }
\end{figure}

The dashed curve in Fig.~5a shows that the flux predicted for a 
spectrum similar to the locally observed
CR protons with the power-law index $\Gamma_{\rm p}=2.75$, fails to
explain
the diffuse $\gamma$-ray flux observed from the inner Galaxy at energies 
above 1\,GeV by a factor of 1.5-2. This deficit cannot be removed by an assumption
of a larger $w_{\rm p}$ or $N_{\rm H}$, because the flux predicted at 500\,MeV 
is already equal to the observed flux. An assumption of a very hard power law
index for the CRs in the Galaxy, $\Gamma_{\rm p}=2.3$ (dot-dashed line), 
explains the data at few GeV, but over-predicts the flux at higher energies.
And finally, a moderately steep 
spectrum of protons with a power-law index    
$\Gamma_{\rm p}\simeq 2.5$ could 
explain the spectral shape of the  observed `GeV bump'  (solid line). 

There is another  possibility  to fit the GeV spectrum of the 
observed diffuse radiation by $\pi^0$-decay $\gamma$-rays.   
In Fig.5b we show the fluxes calculated for the spectra of CR protons  in the form
\begin{equation}
n_{\rm p}(E_{\rm p}) \propto E_{\rm p}^{-\Gamma_{0}}\left[1 +
\left( \frac{E_{\rm p}}{E_{\ast}}\right)^\delta\right]^{-1} \; .
\end{equation}

This spectrum corresponds to the power-law index of protons $ \Gamma_{\rm p}\approx
\Gamma_0$ at energies below some $E_{\ast}$, but $\Gamma_{\rm p}\approx
\Gamma_0 +\delta $ 
at high energies
$E_{\rm p}\gg E_{\ast}$. An energy distribution
of protons of this kind is to be expected if the source function (
acceleration spectrum) of accelerated protons has a single power-law form
with an index $\Gamma_{0}$. Then the energy distribution
of the protons, which practically do not suffer energy losses
(but which nevertheless are formally taken into account in the 
numerical calculations), can be approximated as  
$n_{\rm p}(E_{\rm p}) \simeq q_{\rm p}(E_{\rm p}) \cdot 
\tau_{\rm esc}(E_{\rm p})$. 
This is easily reduced to the form of Eq.(26),
if the diffusive escape time of CRs from the inner Galactic disk 
(Eq.~8) equals the 
convective escape time at $E_{\rm p}=E_{\ast}$.     
The results of calculations in Fig.~5b show that the GeV bump in the spectrum
of diffuse radiation can be well explained (solid line) by a hard spectrum of 
accelerated protons with $\Gamma_{0}\simeq 2.1$ if one takes also into account 
the diffusive escape of particles from the galactic disk 
with a power law index $\delta\sim 0.6$ resulting in a steepening of 
the CR spectrum above the energy $E_{\ast}\sim 10-20\,\rm GeV$.

The results shown in Fig.~5 demonstrate that it is possible 
to explain in a natural way the `GeV' bump observed in the spectrum of diffuse
galactic radiation by $\pi^0$-decay \grs assuming relatively hard 
spectrum of protons at energies  below 100 GeV. 
However,  the best fits in Fig.~5  do not take into account
the fluxes contributed by other \gr production mechanisms, in
particular by the  IC radiation of the electrons. In fact, the IC fluxes
are not negligible and should be taken into account for 
any realistic combination of  parameters characterizing the ISM and 
CRs. It its turn, the  expected IC contribution at GeV energies is  
tightly connected  with the fluxes at lower energies, which are  
contributed  not only by  IC $\gamma$-rays  but  also by  the
bremsstrahlung, and possibly also by the annihilation 
components of radiation.   Therefore,  any attempt to fit the observed 
GeV spectrum of  diffuse radiation by $\pi^0$-decay \grs 
cannot be treated separately,  but rather should  
be conducted  within the multiwavelength approach to the problem. 

\section{Overall fluxes}

All radiation mechanisms considered in previous sections
significantly contribute to the  overall flux of the broad-band 
diffuse $\gamma$-radiation of  the Galactic disk. 
In Fig.~6 we show that the $\gamma$-ray  data from 
$\sim 1\,\rm MeV$ to 30\,GeV can be well explained 
with a set of quite reasonable parameters both for the ISM
and for the acceleration and propagation of
the nucleonic and electronic components of the galactic CRs. 
For calculations in Fig.~6 we have assumed   a mean
line-of-sight depth  for the inner Galactic disk 
$l_{\rm d}=15\,\rm kpc$ as in Fig.~2, 
and a column density $N_{\rm H}=1.5\times 10^{22}\,\rm cm^{-3}$. This 
corresponds to a reasonable mean gas density along the line of sight 
at low galactic latitudes about $n_{\rm H}=N_{\rm H}/l_{\rm d}=
0.33\,\rm cm^{-3}$. Calculations are done so that
the resulting energy distributions of CR protons and electrons 
are normalized to the energy densities $w_{\rm p}=1\,\rm eV/cm^3$
and $w_{\rm e}=0.075\,\rm eV/cm^3$.
The latter value is larger than the energy density of the 
local CR electrons by a factor about 1.5 or so, depending 
on poorly known flux of the local CR electrons below 1 GeV.
Such an enhanced energy density of  
CR electrons  is quite possible if we take into account
that the concentration of CR sources presumably increases towards the
galactic cetre, and that the electrons suffer significant energy losses.  
It is worth notice that the values of the parameters  
$N_{\rm H}$, $l_{\rm d}$, $w_{\rm p}$ and $w_{\rm e}$ may somewhat
vary, but the spectral fits would be essentially
similar to the one in Fig.~6  
if the products  $N_{\rm H}\times w_{\rm p}$  and $n_{\rm H}\times w_{\rm e}$ 
are kept at the same level as in Fig.~6.

\begin{figure}[htbp]
 \resizebox{8.5cm}{!}{\includegraphics{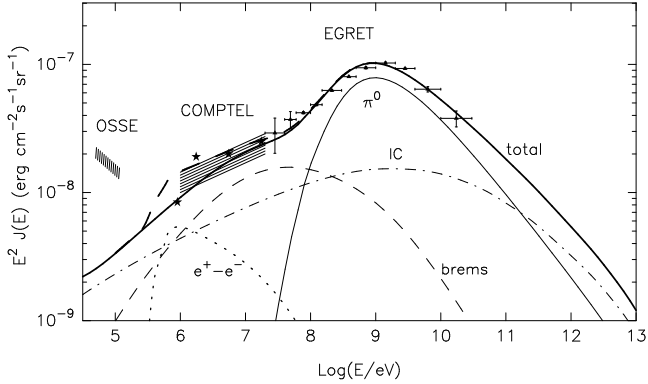}}
\caption{ The fluxes of diffuse radiation produced by both electronic 
and nucleonic components of cosmic rays in the inner Galaxy, calculated  
for a hard power-law source functions of the  electrons with   
$\Gamma_{e,0}=2.15$, and of the protons with   $\Gamma_{\rm p,0}=2.1$,
and
the escape time parameters  $\delta=0.65$, $\tau_{10}=
1.4\times 10^{7}\,\rm yr$, 
$\tau_{\rm conv}=2\times 10^{7}\,\rm yr$. Other model parameters are:
$w_{\rm e}=0.075 \,\rm eV/cm^3$, $w_{\rm p}=1 \,\rm eV/cm^3$, $N_{\rm H}=
1.5 \times 10^{22} \, \rm cm^{-2}$, $l_{\rm d}=15\,\rm kpc$, $B= 6\,\rm
\mu G$. 
Contributions from $\pi^0$-decay (thin solid line), bremsstrahlung 
(dashed), inverse Compton (dot-dashed), and positron annihilation in flight
  (dotted line, for $C_{+}=0.5$) $\gamma$-radiation mechanisms are shown. 
The heavy solid line shows the total flux without contribution from the 
positron annihilation, and the heavy dashed line takes this flux into account. 
 }
\end{figure}

In Fig.~6 hard power law source functions in the form 
of Eq.(11), with $\Gamma_{\rm e,0} =2.15$ and $\Gamma_{\rm p,0}=2.1$
for the electrons and protons respectively, are assumed.    
For the CR escape times in Eqs.~(7) and (8) we have chosen  
$\delta =0.65$, and $\tau_{10} = 1.4\times 10^7\,\rm yr$ and
$\tau_{\rm conv}= 2\times 10^7\,\rm yr$. For these parameters  the diffusive 
escape time at the energy $E_{\ast}=5.8\,\rm GeV$  becomes equal to 
the time of convective escape,   so in the energy range from a few 
GeV to $\sim 10\,\rm GeV$ the spectrum
of CR protons (which do not practically suffer energy
losses)  is  described by Eq.~(26);  it  
gradually steepens from the initial power-law with 
$\Gamma_{\rm p,0}=2.1$ to $\Gamma_{\rm p}=2.75$.  
A decrease of
$E_{\ast}$, as compared with the 
`best fit' value $E_{\ast}= 20\,\rm GeV$ in Fig.~5b, is necessitated by
the increasing contribution of IC radiation to the observed 
diffuse flux at energies $E > 1\,\rm GeV$. Note that 
for CR proton spectra with $\Gamma_{\rm p}=\Gamma_{\rm p,0}+\delta \le 2.6$
this effect of gradually increasing contribution of the IC fluxes at 
GeV energies makes an interpretation of the observational data rather  problematic,
whereas without the contribution of 
the IC component a single power-law spectrum of CR protons
with $\Gamma_{\rm p}\simeq
2.5 $ could explain the `GeV bump' (see Fig.~5a). 
 
The heavy solid curve in Fig.~6 shows the total $\gamma$-ray flux without 
the contribution from relativistic positron annihilation. 
In the energy region below 3 MeV the slope of this curve is noticeably
flatter than the characteristic slope of the fluxes detected 
by COMPTEL (hatched  zone;  Strong et al. 1997). Meanwhile
at MeV energies the flux of this annihilation radiation may significantly
contribute to the overall flux, and it may even exceed the individual 
fluxes of both bremsstrahlung and IC $\gamma$-rays, provided that at
energies below 100 MeV the positron content
in the electronic component of the Galactic cosmic rays is significant.
The dotted
curve in Fig.~6 corresponds to an assumption of a high 
value of $C_{+}=0.5$, and the heavy dashed line shows the 
overall (``IC+bremsstrahlung+annihilation'') flux  
of \grs radiated by the electron component of CRs. For comparison, we show
by stars also the COMPTEL data points (Hunter et al. 1997) corrected
for the contamination caused by the positronium annihilation radiation 
observed (Purcell et al. 1993) in the same direction.
We see that the annihilation of positrons
``in flight'', on top of  the IC and bremsstrahlung fluxes, fits rather well 
the COMPTEL measurements.

Finally, in the context of principal radiation mechanisms of 
the diffuse galactic gamma radiation  at MeV energies, we should mention
also a possible contribution of the prompt $\gamma$-ray line emission
produced by sub-relativistic cosmic rays via nuclear de-excitation.  
The  emissivity of the total (unresolved)  \gr line emission  in the energy 
range between several hundred keV and several MeV, normalized to the
energy density of 
sub-relativistic CRs  $w_{\rm scr}=1 \, \rm eV/cm^3$, and calculated   
for the standard  cosmic composition (of the ambient matter and galactic CRs), 
is about $2 \times 10^{-25}$ 
ph/s H-atom (Ramaty et al. 1979).
This implies that for $N_{\rm H}=1.5 \cdot  10^{22} \, \rm cm^{-2}$ and  
$w_{\rm scr} \leq  \, 1  \,  \rm eV/cm^3$, the energy flux of this  
component of gamma radiation cannot exceed   $10^{-9}$ 
$\rm erg/cm^2 s \, ster$. Therefore,  this radiation mechanism 
could hardly be responsible for more than several per cent of
the observed \gr flux  at MeV energies, unless the energy density of 
sub-relativistic particles in the ISM  significantly exceeds 
the ``nominal'' energy density of CRs in the relativistic regime, 
$w_{\rm scr} \gg w_0 \simeq 1 \, \rm eV/cm^3$.

Hard X-ray emission from the inner parts of the Galaxy has been
recently reported  by  the Ginga and Welcome-1 (Yamasaki et al. 1997) ,
RXTE (Valinia \& Marshall 1998), and  OSEE (Kinzer et al. 1997) teams.   
In Fig.~6 we show the fluxes of diffuse hard X-rays reported 
by OSSE. It is seen that the Bremsstrahlung $+$ IC  
emission of the `radio'  electrons may contribute only $(10-20)\,\%$ to the 
galactic hard X-ray  background.  In order to  explain  the bulk of this  
radiation by the non-thermal bremsstrahlung one has  to postulate an
existence of the interstellar populations of sub-relativistic electrons
(e.g. Yamasaki et al. 1997) and  protons (e.g. Boldt 1999). It should be noted
that the bremsstrahlung of sub-relativistic particles, both of electrons
and protons, is a rather inefficient mechanism of radiation since due to
severe ionization losses  only $\simeq 10^{-5}$  part of the kinetic 
energy of particles is released in the form of  non-thermal hard X-rays
(Aharonian et al. 1979,  Skibo et al. 1996).
Therefore the ``sub-relativistic bremsstrahlung'' models require 
continuous injection of low-energy electrons and/or protons, e.g. by SNRs,  
into the ISM with uncomfortably large rates, especially if one  tries
to explain the fluxes of the low-energy X-rays,  
$W_{\rm scr} \simeq 10^{43}  \, \rm erg/s$ (Skibo et al. 1996).
Moreover, in the case of the proton bremsstrahlung the production of
X-rays is tightly connected with the prompt \gr line emission  
due to the excitation of the nuclei (first of all, Fe, C, O, etc.) of the
ambient gas by the same protons (Aharonian et al. 1979). Therefore this
allows robust upper limits on the flux of the proton
bremsstrahlung X-rays based on the observed fluxes of diffuse $\gamma$-rays 
at MeV energies. The recent analysis by Pohl (1998),  based on 
the comparison of the observed keV and MeV fluxes 
leads to a conclusion that  indeed the  proton bremsstrahlung alone could  
not be responsible for the bulk of the  diffuse Galactic X-ray  emission.     

An alternative mechanism for an explanation of the diffuse X-radiation of
the galactic disc - the  synchrotron emission of ultra-relativistic
electrons in the interstellar magnetic fields - has been suggested 
by Porter \& Protheroe (1997).
An obvious advantage of this mechanism, compared  with the bremsstrahlung
of sub-relativistic  particles, is its almost  $100 \%$   efficiency of
transformation of the kinetic energy of the electrons  
into the X-radiation. On the other hand  this mechanism requires, for any 
reasonably ambient magnetic field, an efficient acceleration of electrons 
up to energies of $300-500 \, \rm TeV$.  
 
These electrons could hardly be produced
by the shocks of SNRs because of severe synchrotron losses on timescale
\begin{equation}
t_{\rm sy}\approx 1.2 \cdot 10^{3} 
\, (E_{\rm e}/100 \, {\rm TeV})^{-1} (B/10 \, {\mu \rm G})^{-2} \, \rm yr.    
\end{equation} 

Comparing this time with the maximum rate of the diffusive
shock acceleration (Lagage \& Cesarsky 1983) 
in the extreme limit of Bohm diffusion,
one easily finds an estimate of the maximum energy of accelerated
electrons
\begin{equation}
E_{0}^{(\rm max )} \simeq 90  \,    \left(\frac{
B}{10 \, \rm {\mu G}}\right)^{-1/2} 
\left(\frac{v_{\rm s}}{4000 \, \rm km/s}\right )
\; \rm TeV \, ,
\end{equation}   
where $v_{\rm s}$ is the shock speed. 
Thus, for typical parameters of a SNR in a stage close to its
Sedov phase, when the acceleration of the bulk of relativistic particles 
takes place (e.g. Berezhko \& V\"olk 1997), namely 
$v_{\rm s} \leq 4000 \, \rm km/s$ and $B \geq 5 \, \rm \mu G$,  the 
characteristic maximum energy of accelerated electrons cannot significantly
exceed 100 TeV. 

More probable sites  for acceleration of electrons to energies
$\gg 100 \, \rm TeV$ could be the shocks terminating  relativistic winds 
driven by pulsars. For example, it is widely believed that the electrons
in the Crab Nebula are accelerated at the termination shock of the
pulsar wind to energies $\sim 10^{15}\,\rm eV$ (see e.g. Arons 1996). 
An assumption that electrons could be accelerated well beyond 100 TeV 
at the wind termination shocks of much older pulsars could then explain the
diffuse hard X-ray background radiation of the Galaxy as a synchrotron emission
of these electrons.  The life time of $E_{\rm e}\sim
300\,\rm TeV$ electrons does not exceed several hundred  years, therefore they 
cannot propagate a distance more (and probably even much less)
than a few tens of parsec from their acceleration sites. It means that 
the  synchrotron origin of the galactic diffuse X-ray background 
would be possible only in a form of  {\it superposition} of the emission 
from a number of unresolved weak and {\it continuous} sources along the line 
of sight. These conditions could be satisfied e.g. by
$10^{5}-10^{6}\,\rm yr$ old pulsars, the overall number of which in the 
inner Galaxy   could be estimated up to 
 $\sim 3\times 10^4$ if neutron stars are
produced with a rate of about 1 per 30 yrs.

\begin{figure}[htbp]
 \resizebox{8.7cm}{!}{\includegraphics{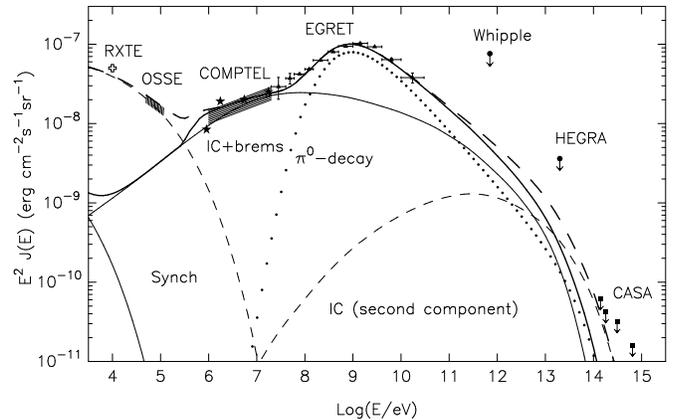}}
\caption{ The diffuse background radiation from the Galactic plane
corresponding to the two-component model for the 
relativistic electrons. The heavy solid line corresponds to the fluxes
produced by the electrons of the first (main) population, with the same
model parameters as in Fig.~6, but for $\delta =0.7$ and $C_{+}=0.3$, and the
heavy dashed line shows the overall fluxes including the contribution
from the second (`pulsar wind'-- see text) population of electrons accelerated
to energies $E_0=250\,\rm TeV$. The local mean magnetic field
for the second electron population is $B_{2}=25\,\rm \mu G$. 
Besides the galactic diffuse background $\gamma$-radiation detected
by COMPTEL and EGRET, the X-ray backgrounds detected by RXTE 
(Valinia \& Marshall 1998) and OSSE (Kinzer et al. 1997), as well as the
upper flux limits at very high energies (VHE) reported by Whipple, HEGRA, 
and CASA-MIA  collaborations are also shown. 
 }
\end{figure}

In Fig.~7 we show the broad band diffuse radiation  flux calculated in the 
framework of the model which assumes that besides the main population of
CRs  (presumably accelerated  by the SNR  shocks), there is also a 
second population of electrons accelerated well beyond 100 TeV at the pulsar
wind termination shocks of old neutron stars/pulsars  in the inner Galaxy. 
The parameters for the first (main) population of accelerated particles, both 
the electrons and protons, are essentially the same as in Fig.~6 (except for 
$\delta=0.7$ 
and $C_{+}= 0.3$ used in Fig.~7).  For the second electron population  
we assume an acceleration spectrum with $\Gamma_{\rm e}=2$ and an exponential 
cutoff energy  $E_{\rm 0}=250\,\rm TeV$, but also a turnover of the 
spectrum at energies
below 1\,TeV.  Note that such a turnover  at low energies is a characteristic 
feature of the electrons accelerated at the pulsar wind termination
shocks (see e.g. Arons 1996), which however does not affect the overall 
flux of diffuse $\gamma$-rays below TeV energies (dominated by the radiation
of the main component of CRs). For the mean magnetic field of the ISM 
in Fig.~7 we assume  the same value as in Fig.~6,  
$B=B_1=6\,\rm \mu G$.  But for the second electron component 
we assume significantly larger ambient field, $B_2=25\,\rm \mu G$,
in order to fit the hard X-ray data. Different 
magnetic fields for the first and second components of the 
synchrotron radiation  can be understood if we remember that the multi-TeV 
electrons cool very rapidly therefore they cannot propagate very far from their 
accelerators. More specifically,  the hypothesis that the {\em second} 
(ultra-high energy) population of electrons   is produced at the termination shocks of 
relativistic pulsar winds (which inject not only relativistic electrons, but also magnetic fields)
could explain also why the magnetic field $B_2$ could be
much higher than the mean field $B_1$ in the Galaxy.  

The assumption of high magnetic field $B_2$ becomes very
important for  a self-consistent interpretation of the `diffuse' flux of
hard X-rays detected in the galactic plane in terms of synchrotron 
radiation of $E_{\rm e} \geq 300\,\rm TeV$ electrons, because otherwise 
the fluxes of IC radiation produced by the 
same electrons would exceed the upper flux limits reported by CASA-MIA
collaboration  at energies $E_{\gamma} \geq 100\,\rm TeV$ (Barione et al. 
1998). As it is seen in Fig.~7, the  energy flux of OSSE exceeds the
flux upper limits of CASA by a factor of $\geq 300$. The ratio of
energy fluxes of synchrotron to IC radiations
produced by an electron is about the ratio of the magnetic field to the
soft photon field energy densities, i.e. 
$(B^2/8\pi)/0.25\,\rm eV/cm^3 = 10 (B/10\,\rm \mu G)^2$, if the
IC scattering occurs in the Thomson limit, when the parameter
$b=4 \epsilon_0 E_{\rm e}/(m_{\rm e} c^2)^2 \ll 1$ 
($\epsilon_0\sim 6.5\times 10^{-4}\,\rm eV$ is the mean energy of 2.7\,K
MBR photons).  Actually, the IC scattering of $E_{\rm e} \sim 300\,\rm
TeV$ electrons occurs in a moderate
Klein-Nishina regime with  $b\sim 3$. For these values of
parameter $b$ the emissivity of IC radiation is reduced (compared with the 
Thomson limit) by a factor 5-10, therefore the ratio of the synchrotron to
IC fluxes $\geq 300$ implies a magnetic field $B_2\geq 20\,\rm \mu G$.    

The acceleration power in the second electron component which is needed
for interpretation of the hard X-ray background in Fig.~7  
is about $6\times 10^{36}\,\rm erg/s$ per $\rm kpc^3$, or about
$L^{\rm (II)}_{\rm e} \simeq 1.4 \times 10^{39}\,\rm erg/s$ in the entire
inner galactic disk with a radius about 8\,kpc   
and a thickness $h\simeq 1\,\rm kpc$.  
 For the overall number of such pulsars of order $3\times 10^4$,
this implies a rather modest  mean acceleration  power per 1 `old pulsar' 
of about $5\times 10^{34}\,\rm erg/s$, i.e. 
by four orders of magnitude less than the power of the 
relativistic electron-positron wind of the  Crab pulsar.  
Note that the kick velocities 
of the pulsars can be of order from a few 100 to $\sim 1000\,\rm km/s$,
so the $10^6\,\rm yr$ old pulsars would be able to propagate to distances 
$\leq (0.3-1) \,\rm kpc$, contributing therefore to the emission at the  
galactic latitudes up to several degree. Note also that even for $10^4$
pulsars in the $10^\circ \times 90^{\circ}$ field of view of the inner
Galaxy considered here, the mean density of such pulsars on the sky 
corresponds to $\simeq 10$ per square degree. Therefore in principle for
the hard X-ray/soft $\gamma$-ray detectors, which typically do not have an
excellent angular resolution, a superposition of the emission of large 
number of such relatively weak sources (with sizes presumably $\geq 10
\,\rm pc$ ) in the field of view of the detector could imitate   a
"diffuse" emission.   

\section{Discussion}

The diffuse galactic gamma ray emission 
carries a unique information, a proper
understanding of which would eventually result in a quantitative 
theory of the origin of galactic cosmic rays. Although the 
problem is essentially complicated and confused because of
several competing mechanisms of production of diffuse $\gamma$-rays, 
the data obtained by the COMPTEL and 
EGRET  detectors aboard Compton GRO  allow rather definite conclusions
concerning  the relative contributions of different \gr production mechanisms in the
energy region from 1 MeV to  30 GeV.  

At energies below 100 MeV the diffuse $\gamma$-radiation
has an electronic origin. For a set of reasonable parameters for both 
the ISM and CR electrons,  the observed \gr fluxes from 10 MeV to 100 MeV 
can be well explained by  the superposition of the bremsstrahlung and the 
IC components of radiation of relativistic electrons. At lower energies, a non-negligible
flux can be  contributed by the annihilation of relativistic positrons 
with the ambient thermal electrons. Moreover, if  the content of positrons
at energies 
$E \leq 10 \, \rm MeV$ is close to  $e^{+}/(e^{-}+e^{+}) \sim 0.5$, the
\gr ray flux around 1 MeV would be  dominated by the non-thermal annihilation
radiation. Between several MeV and 100 MeV the radiation is dominated by 
the electron  bremsstrahlung.  This result agrees with previous studies 
(Kniffen \& Fichtel 1981,  Sacher \& Sch\"onfelder  1984, Gehrels \&
Tueller 1993). 

Above 100 MeV the IC process  
dominates in the production of diffuse 
$\gamma$-rays,
but at these energies a new mechanism is needed 
in order to explain the so called ``GeV bump''  detected by EGRET.  
We believe that this distinct  emission feature 
of the inner Galaxy could be naturally explained by  \grs of nucleonic
origin produced at the interactions of CR protons and nuclei with the
ambient interstellar
gas. Indeed, the overall  observed MeV/GeV emission of the inner Galaxy  
is well fitted by the $\pi^0$-decay  \grs 
(on top of ``IC + bremsstrahlung + annihilation'' contribution by CR electrons) 
provided that the spectrum of CR protons in the inner galactic disk  at 
low energies is substantially  flatter than the spectrum of directly observed 
local CRs. In particular, a proton spectrum in the form of Eq.(26) with  
$E_\ast\sim 10 \, \rm GeV$, $\Gamma_0=2.1$ and $\delta\simeq 0.6$, which
could be formed 
due to a reasonable combination of diffusive and convective escape time-scales 
of CRs from the inner galactic disk, is able to explain very well
the observed diffuse galactic \gr spectrum (see Fig. ~6 and Fig.~7) .
It is interesting to note that in the total ``$\pi^0$+IC'' spectrum
the contribution of $\pi^0$-decay \gr component  
gradually  decreases since the spectrum of CR  protons at high energies
with an index $\Gamma_{\rm p} \approx \Gamma_0 + \delta \simeq 2.7$
is steeper than the spectrum of the diffuse gamma radiation 
with a power-law index $\simeq 2.5$ observed at  energies above several
GeV.  
However, this decline is compensated by the hard IC component of 
radiation which at energies above 30 GeV becomes the dominant contributor
to the overall \gr flux. For convenience of further discussion, we will
call this possibility for explanation of the GeV bump in the observed 
\gr spectrum the scenario 1. 

     Another possibility for explanation of the spectrum of $\gamma$-rays
up to 30 GeV could be  a single power-law spectrum of CR protons
with $\Gamma_{\rm p}\sim 2.5$, but then we should assume  
a significant reduction of the  IC contribution to the overall \gr flux 
at such high energies. This possibility, which we call the scenario 2, 
can be realized if the  acceleration spectrum of CR electrons does not 
extend to TeV energies.  In this scenario
the \gr fluxes below several GeV are explained, 
as in the scenario 1,  by superposition of the electronic and nucleonic
components of radiation.   However at higher 
energies the \gr flux would be strongly dominated by $\pi^0$-decay component of
radiation.  For the  acceleration spectrum of protons  with  
$\Gamma_0\sim 2.1$, the required  single power-law spectrum  of protons in the 
inner Galaxy  with $\Gamma_{\rm p} \sim 2.5$ could be formed if
$\delta\sim 0.4$ and  $E_\ast \ll 10 \, \rm GeV$ (see Eq.~26).
This conclusion would imply  that the mean time of convective 
escape of CRs  from the disk is much larger 
than the  diffusive escape of particles  at 10 GeV.

These two scenarios predict essentially different origin of $\gamma$-rays
in the VHE domain. 
While in the  scenario 1 the \gr background of the galactic disk at 
$E \geq 100 \, \rm GeV$ is contributed mainly by  
IC scattering of multi TeV electrons on 2.7 K MBR,  in the  scenario 2  the 
VHE \gr flux is  dominated by $\pi^0$-decay $\gamma$-rays. 
Therefore the future spectroscopic  measurements of the diffuse radiation of 
the inner Galaxy  at low ($|b| \leq 2^{\circ}$) and high  (e.g. $2^{\circ} \leq |b| \leq 10^{\circ}$)  
galactic latitudes by GLAST  at $E \leq 100 \, \rm GeV$, 
and by  imaging atmospheric Cherenkov telescope arrays H.E.S.S. 
and CANGAROO-3  (both to be located in  the Southern Hemisphere)
above 100 GeV could  provide crucial information about the character  
of propagation of   TeV cosmic rays in the galactic disk (Weekes et al. 1997).

Because all target photon fields for production of IC \grs extend 
practically  with the same intensity (see e.g. Chi \& Wolfendale 1991)
well above the characteristic height $h \sim 200 \, \rm pc$ of the
gaseous disk,  and since the fluxes of very high energy electrons  above 
the galactic plane  up to several hundred  pc  could be still significant,  
the realization of the scenario 1 can be distinguished from the 
scenario 2 by  different spectra  of high energy 
radiation detected at different galactic latitudes $|b|$.   In particular,  the 
scenario 1 predicts  an increase of the  relative contribution 
of the IC component to 
the \gr flux, therefore we should  expect some flattening
(depending on the rate of decline of VHE electron fluxes above the galactic plane) 
of the spectrum of \grs at $|b|$ larger than few degree.  Note, however, that 
the measurements  at different $|b|$ could not distinguish between the scenario
2, which interprets the observed Galactic `excess' GeV radiation 
in terms of truly diffuse emission of (mainly) nucleonic origin
from the models interpreting this  radiation as  a superposition 
of contributions from unresolved  SNRs 
(V\"olk 1999,  Berezhko \& V\"olk 2000), or ``active'' molecular clouds 
(Aharonian 1991, Aharonian \& Atoyan 1996). 
The level of contamination of the  {\it truly diffuse}  radiation of the disk 
by faint, but numerous \gr sources could be properly
estimated only by  future measurements  of the angular 
distribution of  the galactic  background in 
scales less than  $1^{\circ}$. The GLAST with its angular  resolution 
as good as $0.1^{\circ}$ and detection area 
by a factor of 10 larger than EGRET (see e.g. Bloom 1996),  
is nicely suited  for this task. 

The fluxes of \grs expected above 1 TeV in both scenarios 1 and 2   
are below the current flux upper limits set by the Whipple 
(Reynolds et al. 1993),  HEGRA (Schmele 1998) and CASA-MIA groups 
(see Fig. 8). However larger fluxes at such high energies cannot be excluded. 
Indeed, besides the hard $\pi^0$-decay  \gr component 
due to CR sources which could remain unresolved, we may expect also 
higher IC fluxes of \grs radiated by  a 
possible second  component of ultra  high energy electrons with a
spectrum 
extending well beyond 100 TeV. The existence of such a component of 
electrons is needed within the model which interprets the  hard X-radiation 
of the inner galactic disk  as a result of synchrotron radiation of 
ultra-relativistic electrons. This hypothesis requires a very large  (up
to $E_{\rm max} \sim 10^{15} \, \rm eV$) 
maximum energy of  electrons and  large magnetic field $B_2 \geq 20 \, \rm \mu G$  
in order to fit the observed X-ray fluxes, as well as to  
avoid an overproduction of IC \grs  at 100 TeV.  Such large values of both  
$E_{\rm max}$ and $B_2$ could be probably explained assuming that
this energetic component of electrons is produced at the wind termination
shocks 
of the ensemble of old pulsars/neutron stars in the galactic disk. 
The hypothesis of the synchrotron origin of hard diffuse X-radiation 
of the galactic disk  needs further confirmation based on the detailed
study of spectral and spatial distribution of hard X-rays by future satellite
missions like ASTRO-E and INTEGRAL, as well as by a detection of 
the diffuse \gr background of the galactic disk at  $E \sim  100 \, \rm
TeV$ 
at the flux level rather close to   the  current CASA-MIA  upper limits.

\begin{acknowledgements}
We thank the anonymous referee for his very helpful
and important comments. The work of AMA has been partly supported
through the Verbundfoeschung Astrono-mie/Astrophysik of the German
BMBF under grant No. 05-2HD66A(7).   
\end{acknowledgements}

\end{document}